\RequirePackage{fix-cm}
\documentclass[smallcondensed]{svjour3}
\usepackage{graphicx}
\usepackage{caption}
\usepackage{amsmath}
\usepackage{amssymb}

\usepackage{isabelle,isabellesym}
\usepackage{url}

\usepackage{algorithm}
\usepackage{algpseudocode}

\usepackage{booktabs}

\usepackage{tikz}
\tikzset{every picture/.style={remember picture}}

\usepackage{hyperref}

\newcommand{\extR}{\overline{\mathbb{R}}}
\newcommand{\TaQ}{\mathrm{TaQ}}
\newcommand{\SRemS}{\mathrm{SRemS}}
\newcommand{\SPRemS}{\mathrm{SPRemS}}

\newcommand{\Var}{\mathrm{Var}}

\renewcommand{\div}{\mathbin{\mathrm{div}}}
\renewcommand{\mod}{\mathbin{\mathrm{mod}}}
\renewcommand{\pmod}{\mathbin{\mathrm{pmod}}}
\newcommand{\pdiv}{\mathbin{\mathrm{pdiv}}}

\newcommand{\sgn}{\mathrm{sgn}}

\newcommand{\jump}{\mathrm{jump}}

\begin{document}

\title{Deciding Univariate Polynomial Problems Using Untrusted Certificates in Isabelle/HOL
\thanks{
        The first author was funded by the China Scholarship Council, via the CSC Cambridge Scholarship programme.
        The development of MetiTarski was supported by the Engineering and Physical Sciences
        Research Council [grant numbers EP/I011005/1, EP/I010335/1].}
}

\titlerunning{Univariate Decision Procedure}

\author{Wenda Li  \and Grant Olney Passmore \and Lawrence C. Paulson}

\institute{Wenda Li \at
        Computer Laboratory, University of Cambridge \\
        \email{wl302@cam.ac.uk}           %  \\
        \and
        Grant Olney Passmore \at
        Aesthetic Integration, London and Clare Hall, University of Cambridge \\
        \email{grant.passmore@cl.cam.ac.uk}
        \and
        Lawrence C. Paulson \at
        Computer Laboratory, University of Cambridge \\
        \email{lp15@cam.ac.uk}           %  \\
}

%\institute{Computer Laboratory, University of Cambridge\\
%\email{\{wl302,lp15\}@cam.ac.uk}
%\and
%Aesthetic Integration, London and Clare Hall, University of Cambridge
%\email{grant.passmore@cl.cam.ac.uk}
%}

\maketitle

\begin{abstract}%HOL\@.
We present a proof procedure for univariate real polynomial problems
in Isabelle/HOL. The core mathematics of our procedure is based on
univariate cylindrical algebraic decomposition. We follow the approach
of untrusted certificates, separating solving from verifying:
efficient external tools perform expensive real algebraic
computations, producing evidence that is formally checked within
Isabelle's logic. This allows us to exploit highly-tuned computer
algebra systems like Mathematica to guide our procedure without
impacting the correctness of its results.
We present experiments demonstrating the efficacy of this approach, in
many cases yielding orders of magnitude improvements over previous
methods.
\keywords{\and Interactive theorem proving \and Isabelle/HOL \and Decision procedure \and Cylindrical algebraic decomposition}
\end{abstract}

\section{Introduction}
Nonlinear polynomial systems are ubiquitous in science and
engineering. As real-world applications of formal verification
continue to grow and diversify, there is an increasing need for proof
assistants (e.g., ACL2, Coq, Isabelle \cite{isa-tutorial}, HOL
Light and PVS) to provide automation for reasoning about nonlinear systems
over the reals
\cite{narkawicz2015formally,daumas2009verified,pvs_bernstein}.

Cylindrical algebraic decomposition (CAD) \cite{collins1976quantifier} is one of
the most powerful known techniques for analysing non-linear polynomial systems.
CAD-based methods have been implemented in various systems such as Z3
\cite{de2008z3}, QEPCAD \cite{brown2003qepcad}, Mathematica and Maple. However,
implementing CAD-based decision procedures within proof assistants has been
hindered by the difficulty in formalising the mathematics justifying CAD
computations.

In this paper, we present a formally verified
procedure{\footnote{Code is available from
  \url{https://bitbucket.org/liwenda1990/src_jar_2017}}}
based on CAD for univariate polynomial problems with rational
coefficients. Goals such as
\[
 \forall x.\, (x^2>2 \wedge x^{10}-2x^5+1 \geq 0) \vee x<2
\]
\[
 \exists x.\, (x^2=2 \wedge (x>1 \vee x<0))
\]
can be discharged by our tactic automatically. It should be noted that
certifying a general \emph{multivariate} CAD procedure is much harder,
and the univariate version we describe in the paper is only a first
step in that direction.

A key feature of our procedure is its certificate-based design in which an
external untrusted (but ideally highly efficient) program is used to find
certificates, and those certificates are then checked by verified internal
procedures. Overall, the soundness of our procedure depends solely on the
soundness of Isabelle's logic (and code generation\footnote{As our tactic is
  computationally intense, our procedure makes use of the proof by reflection
  technique \cite{haftmann2010code}.}) rather than trusted external oracles.
This is much like Isabelle's sledgehammer tactic, which sceptically incorporates
various external tools.

Our main contributions are:
\begin{itemize}
\item an efficient formalised theory of Tarski queries,
\item an efficient approach to univariate sign determination at real algebraic points,
\item a practical formally verified procedure for
  real algebraic problems based on univariate CAD.
\end{itemize}

The paper continues at follows: A motivating example
(\S\ref{sec:motivating_ex}) and a description of the overall design
(\S\ref{sec:overall_design}) sketch the general idea of our
procedure. The construction and manipulation of real algebraic numbers
is developed in (\S\ref{sec:encoding_algebraic}), including a sign
determination procedure for evaluating polynomials at real algebraic
points (\S\ref{sec:deciding_sign}). The main proof is described in
(\S\ref{sec:main_proof}), which is followed by a discussion of
interaction with external solvers
(\S\ref{sec:linking_to_external}). Next, experiments and related work
(\S\ref{sec:related_work}) are described along with further discussion
of our tactic (\S\ref{sec:discussion_metitarski}). We then conclude
with a look towards the future (\S\ref{sec:conclusion}).

\section{A Motivating Example} \label{sec:motivating_ex}

\begin{figure} [h]
        \centering
        \begin{tikzpicture}[scale=0.82]
        \draw[->] (-4,0) -- (5,0) node[right] {$x$};
        %\draw[->] (0,-3) -- (0,4.2) node[above] {$y$};
        \draw[domain=-3.5:3.5,smooth,variable=\x,blue] plot ({\x},{0.5*\x*\x-1});
        \draw[domain=-3.5:3.5,smooth,variable=\x,red]  plot ({\x},{\x + 3});

        \draw[fill=black] (-3,0) circle (0.05) node [above left] {$-3$};
        \draw[fill=black] (-1.414,0) circle (0.05) node [above right] {$-\sqrt{2}$};
        \draw[fill=black] (1.414,0) circle (0.05) node [above left] {$\sqrt{2}$};

        \draw[dashed] (-3,-1) -- (-3,5);
        \draw[dashed] (-1.414,-1) -- (-1.414,5);
        \draw[dashed] (1.414,-1) -- (1.414,5);

        \draw (3.3,4) node[right] {$P(x)=\frac{1}{2}x^2-1$};
        \draw (3.7,6.5) node[right] {$Q(x)=x+3$};

        %\draw[xshift=-2cm,yshift=-5cm,-] (-4,0) -- (-3,0) node[right] {};
        %\draw[xshift=-2cm,yshift=-5cm, domain=-4:-3,smooth,variable=\x,blue] plot ({\x},{0.25*\x*\x-0.5});
        %\draw[xshift=-2cm,yshift=-5cm, domain=-4:-3,smooth,variable=\x,red]  plot ({\x},{0.5*\x + 1.5});
        %\draw[xshift=-2cm,yshift=-5cm] (-3,0) circle (0.05);
        %
        %
        %\draw[xshift=-1cm,yshift=-5cm,fill=red,red] (-3,0) circle (0.05);
        %\draw[xshift=-1cm,yshift=-5cm,fill=blue,blue] (-3,1.75) circle (0.05);

        \end{tikzpicture}
        \caption{The plot of $P(x)=\frac{1}{2}x^2-1$ and $Q(x)=x+3$} \label{fig:univ_cad}
\end{figure}
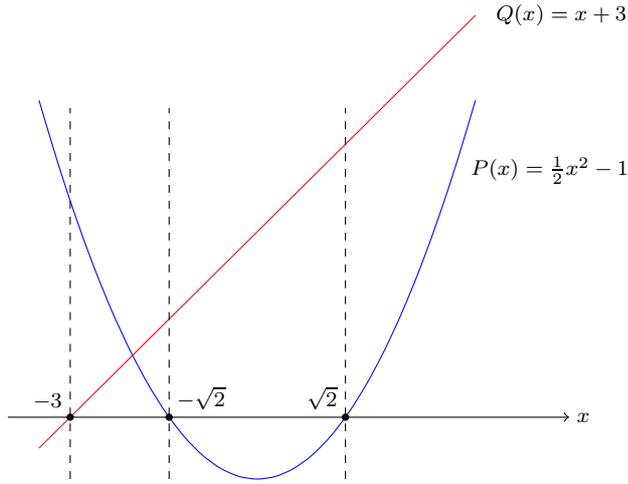

Unlike the general case of $\mathbb{R}^n$, the restriction of CAD to
univariate problems (i.e., to $\mathbb{R}^1$) is relatively
straight-forward.
Suppose we wish to prove
\[
        \forall x.\, P(x)>0 \vee Q(x) \geq 0
\]
where
\[
        P(x)=\frac{1}{2}x^2-1
\]
\[
        Q(x)=x+3.
\]

To do so, we can \emph{decompose} $\mathbb{R}$ into disjoint connected
components induced by the roots of $P$ and $Q$.
This is illustrated in Fig. \ref{fig:univ_cad}:

\vbox{
\[
\mathfrak{D} =\{(-\infty,-3),\{
\tikz[anchor=base, baseline]{\node(q1){$-3$};}
\},(-3,-\sqrt{2}) , \{
\tikz[anchor=base, baseline]{\node(p1){$-\sqrt{2}$};}
\} ,(-\sqrt{2},\sqrt{2}), \{
\tikz[anchor=base, baseline]{\node(p2){$\sqrt{2}$};}
\}
, (\sqrt{2},\infty)  \}
\]
\vspace{0.5cm}\par
% insert text
\noindent\hfil
\begin{tikzpicture}
\node  (qr) {root of $Q$};

\node  [right,text width=2.5cm,align=center] (pr) at (qr.east){roots of $P$};
%
%\node[above right] (t2) at (t1.east) {previous estimate};
%\node[right,text width=2.5cm,align=center] (t3) at (t2.east)
%{(gain)\\ The weight\\ of the adjustment};
%\node[right,text width=2.5cm,align=center] (t4) at (t3.east)
%{(gain)\\ The~weight\\ of the adjustment};
\end{tikzpicture}
\begin{tikzpicture}[overlay]
\draw[blue,thick,->] (q1) to [in=90,out=265] (qr.north);
\draw[blue,thick,->] (p1) to [in=90,out=265] (pr.north);
\draw[blue,thick,->] (p2) to [in=60,out=235] (pr.north);
\end{tikzpicture}
}

\noindent and it can be observed that both $P$ and $Q$ have invariant
signs over each of these components. For example, as can be seen from
Fig.  \ref{fig:univ_cad}, $P(x)<0$ and $Q(x)>0$ hold for all $x \in
(-\sqrt{2},\sqrt{2})$. To decide the conjecture, we can pick sample
points from each of these components and evaluate $\lambda x.\, P(x) >
0 \vee Q(x) \geq 0$ at these points. That is,

\begin{equation} \label{eq:forall_ex}
\begin{split}
&\forall x.\, P(x)>0 \vee Q(x) \geq 0 \\
&= \forall D \in \mathfrak{D}.\, \forall x \in D.\, P(x)>0 \vee Q(x) \geq 0 \\
&= \forall x \in \{-4,-3,-2,-\sqrt{2}, 0, \sqrt{2}, 2 \}.\, P(x)>0 \vee Q(x) \geq 0 \\
&= (P(-4)>0 \vee Q(-4) \geq 0) \wedge (P(-3)>0 \vee Q(-3) \geq 0) \wedge \dots \\
&\qquad	\wedge (P(2)>0 \vee Q(2) \geq 0) \\
&= \mathrm{True} \\
\end{split}
\end{equation}
since
\[
\begin{split}
-4 &\in  (-\infty,-3) \\
-3 &\in \{-3\} \\
-2 & \in (-3,-\sqrt{2}) \\
-\sqrt{2} &\in \{-\sqrt{2}\} \\
0 &\in (-\sqrt{2},\sqrt{2})\\
\sqrt{2}&\in \{\sqrt{2}\}\\
2 &\in (\sqrt{2},\infty).  \\
\end{split}
\]

Analogously, to decide an existential formula
\[
\exists x.\, P(x) =0 \wedge Q(x) >0,
\]
we have
\begin{equation} \label{eq:exists_ex}
\begin{split}
&\exists x.\, P(x)=0 \wedge Q(x) > 0 \\
&= \exists D \in \mathfrak{D}.\, \exists x \in D.\, P(x)=0 \wedge Q(x) > 0 \\
&= \exists x \in \{-4,-3,-2,-\sqrt{2}, 0, \sqrt{2}, 2 \}.\, P(x)=0 \wedge Q(x) > 0 \\
&= (P(-4)=0 \wedge Q(-4) > 0) \vee (P(-3)=0 \wedge Q(-3) > 0) \vee \dots \\
&\qquad	\vee (P(2)=0 \wedge Q(2) > 0) \\
&= \mathrm{True}. \\
\end{split}
\end{equation}

In performing these arguments, there were a few ``obvious'' subtleties:

\begin{itemize}
\item The decomposition of $\mathbb{R}$ into the seven regions given
  \emph{covered} the entire real line. That is,
  \[ (-\infty,-3) \ \cup \ \{-3\} \ \cup \ (-3,-\sqrt{2}) \ \cup \ \{-\sqrt{2}\}
    \ \cup \ (-\sqrt{2},\sqrt{2}) \ \cup \ \{\sqrt{2}\} \ \cup \
    (\sqrt{2},\infty) \ = \ \mathbb{R}. \]

\item The ``sign-invariance'' of $P$ and $Q$ over each region was
  exploited to allow only a single sample point to be selected from
  each region. This property holds as by the Intermediate Value
  Theorem, $P$ and $Q$ can only change sign by passing through a root.
\item The signs of univariate polynomials were evaluated at irrational real
  algebraic points like $\sqrt{2}$ to determine the truth values of atomic formulas.
\end{itemize}

In creating our automatic proof procedure, all of this routine reasoning must, of
course, be formalised.
Moreover, the isolation of polynomial roots (and thus sign-invariant regions)
and the sign determination for polynomials at real algebraic points are
computationally expensive operations.
Computer algebra systems like Mathematica have decades of tuning in their
implementations of these core algebraic algorithms.
To have a practical proof procedure, we wish to take advantage of these highly
tuned external tools as much as possible.
Let us next describe how this can be done.

\section {A Sketch of our Certificate-based Design} \label{sec:overall_design}

There is a rich history of certificate-based, sceptical integrations between
proof assistants and external solvers.
Examples include John Harrison's sums-of-squares method \cite{daumas2009verified}
and the Sledgehammer \cite{paulson2010three} command in Isabelle.
%
% TODO(g): Add John Harrison's differentiation/integration link from the 90s (with Maple?)

Certificate-based approaches are motivated by many observations, including:
\begin{itemize}
\item External solvers are often highly tuned and run much faster than verified ones.
\item Verification of certificates from external solvers is usually much easier
  than finding them. Such verification ensures the soundness of the overall tactic.
\item Switching between different external solvers does not require changes in formal proofs.
\end{itemize}

\begin{algorithm}[h]
\caption{Prove univariate universal formulas over reals}\label{alg:universal}
\begin{algorithmic}[1]
\Require $F(x)$ is a quantifier-free formula over reals
\Ensure Return true if $\forall x.\, F(x)$ holds
\Procedure{universal}{$\forall x.\, F(x)$}
   \State $\mathfrak{P}\gets$ extract polynomials from $F(x)$ \Comment{$\mathfrak{P} \subseteq \mathbb{Z}[X]$}
        \State {\color{red} $\mathit{roots}\gets$ real roots of $\mathfrak{P}$} \Comment{Roots returned by external programs}
   \State $\mathit{samples} \gets$ construct sample points from $\mathit{roots}$

   \If{$(\forall x \in \mathit{samples}.\, F(x)) \wedge (\mathit{roots} \text{ are indeed all real roots of } \mathfrak{P})$}
   \State \textbf{return} true
   \EndIf
\EndProcedure
\end{algorithmic}
\end{algorithm}

Algorithm~\ref{alg:universal} sketches our idea for univariate universal
formulas. In particular, in line 3, we use external programs to return real
roots of polynomials (i.e., $\mathfrak{P}$) from the quantifier-free part of the
formula (i.e., $F(x)$). Those roots (i.e., $\mathit{roots}$) correspond to a
decomposition such that each polynomial from $\mathfrak{P}$ has a constant sign
over each component of this decomposition. Since the roots are returned by
untrusted programs, in line 5, we not only check $\forall x \in
\mathit{samples}.\, F(x)$ as in Equation (\ref{eq:forall_ex}) but also certify
that these roots are indeed all real roots of $\mathfrak{P}$.

The step in line 3 in Algorithm~\ref{alg:universal} is more commonly
referred as \emph{(real) root isolation}, which is a classic and
well-studied topic in symbolic computing. Although we can in principle
formalise our own root isolation procedure (e.g., using the
Sturm-Tarski theorem), it is utterly unlikely that our implementation
will be competitive with state-of-the-art ones, especially for
polynomials of high degree, large bit-width, or whose roots are very
close together. Therefore, we delegate this computationally expensive
step to external tools.

\begin{algorithm}[h]
        \caption{Prove univariate existential formulas over reals}\label{alg:existential}
        \begin{algorithmic}[1]
                \Require $F(x)$ is a quantifier-free formula over reals
                \Ensure Return true if $\exists x.\, F(x)$ holds
                \Procedure{existential}{$\exists x.\, F(x)$}
%		\hspace*{-\fboxsep}\colorbox{green}{\parbox{\linewidth}{\STATE $bar(elem)$}}
                \State {\color{red} $r \gets$  solution to $F(x)$} \Comment{Solution returned by external programs}
                \If{$F(r)$}
                \State \textbf{return} true
                \EndIf
                \EndProcedure
        \end{algorithmic}
\end{algorithm}

With existential formulas, the situation is even simpler as illustrated in
Algorithm~\ref{alg:existential}, since we do not need to deal with the
decomposition internally. Rather, all we need is a real algebraic witness that
satisfies $\lambda x.\, F(x)$ to certify $\exists x.\, F(x)$. What is more
interesting is that the satisfaction problem for $\lambda x.\, F(x)$ can be not
only solved by a CAD procedure, which is complete but not very fast due to its
symbolic nature, but also be complemented by
highly efficient incomplete numerical methods. Thus it is natural to externalize
the step in line 2 in Algorithm~\ref{alg:existential}.

\section{Encoding Real Algebraic Numbers} \label{sec:encoding_algebraic}
External programs in either Algorithm~\ref{alg:universal} and
\ref{alg:existential} can return real algebraic numbers (e.g. $\sqrt{2}$). In
this section, we see how to formalise such numbers in Isabelle/HOL.

The real algebraic numbers ($\mathbb{R}_{\mathrm{alg}}$) are real roots of non-zero
polynomials with integer (equivalently, rational) coefficients. They form a
countable, computable subfield of the real numbers. To encode them, we use a
polynomial with integer coefficients and a root selection method to ``pin down''
the root in question. Common root selection methods include isolating intervals,
root indices or Thom encodings. We use the root interval approach, that is, a
real algebraic number $r \in \mathbb{R}_{\mathrm{alg}}$ will be given by
\begin{itemize}
\item a polynomial $p \in \mathbb{Z}[x]$ s.t. $p(r) = 0$, and
\item two rationals $a,b \in \mathbb{Q}$ s.t. $r$ is the only root of $p$
  contained in $[a,b]$.
\end{itemize}

To reason over the reals, we define a function \isa{Alg} to embed those real
algebraic numbers into the reals:
\begin{isabelle}
        Alg{\isacharcolon}{\isacharcolon}\ {\isachardoublequoteopen}int\ poly\ {\isasymRightarrow}\ float\ {\isasymRightarrow}\ float\ {\isasymRightarrow}\ real{\isachardoublequoteclose}
\end{isabelle}
where \isa{int poly} is a polynomial with integer coefficients and the two
\isa{float} arguments represent an interval. Note, a \isa{float} in Isabelle/HOL
is a dyadic rational number of the form
\[
        a 2^b \quad \text{where} \quad a,b \in \mathbb{Z}.
\]
Compared to our previous work \cite{li2016modular}, where a pair of rational
numbers is used to represent an interval, the dyadic rational approach is more
efficient due to the elimination of ubiquitous greatest common divisor (gcd)
operations within rational arithmetic.

In Isabelle/HOL, a real number is represented as a Cauchy sequence of type
\isa{nat\ \isasymRightarrow\ rat}, where a Cauchy sequence is defined as
\begin{isabelle}
        \isacommand{definition}\isamarkupfalse%
        \isanewline
        \ \ cauchy\ {\isacharcolon}{\isacharcolon}\ {\isachardoublequoteopen}{\isacharparenleft}nat\ {\isasymRightarrow}\ rat{\isacharparenright}\ {\isasymRightarrow}\ bool{\isachardoublequoteclose}\isanewline
        \isakeyword{where}\isanewline
        \ \ {\isachardoublequoteopen}cauchy\ X\ {\isasymlongleftrightarrow}\ {\isacharparenleft}{\isasymforall}r{\isachargreater}{\isadigit{0}}{\isachardot}\ {\isasymexists}k{\isachardot}\ {\isasymforall}m{\isasymge}k{\isachardot}\ {\isasymforall}n{\isasymge}k{\isachardot}\ {\isasymbar}X\ m\ {\isacharminus}\ X\ n{\isasymbar}\ {\isacharless}\ r{\isacharparenright}{\isachardoublequoteclose}
\end{isabelle}
We then convert an encoding of a real algebraic number into a sequence of type
\isa{nat\ \isasymRightarrow\ rat}. The idea is to bisect the isolating interval
through each recursive call, and proceed with the half where the sign of the
polynomial changes at its end points:
\begin{isabelle}
        \isacommand{fun}\isamarkupfalse%
        \ to{\isacharunderscore}cauchy{\isacharcolon}{\isacharcolon}\ {\isachardoublequoteopen}rat\ poly\ {\isasymtimes}\ rat\ {\isasymtimes}\ rat\ {\isasymRightarrow}\ nat\ {\isasymRightarrow}\ rat{\isachardoublequoteclose}\ \isakeyword{where}\isanewline
        \ \ {\isachardoublequoteopen}to{\isacharunderscore}cauchy\ {\isacharparenleft}{\isacharunderscore}{\isacharcomma}\ lb{\isacharcomma}\ ub{\isacharparenright}\ {\isadigit{0}}\ {\isacharequal}\ {\isacharparenleft}lb{\isacharplus}ub{\isacharparenright}{\isacharslash}{\isadigit{2}}{\isachardoublequoteclose}{\isacharbar}\isanewline
        \ \ {\isachardoublequoteopen}to{\isacharunderscore}cauchy\ {\isacharparenleft}p{\isacharcomma}\ lb{\isacharcomma}\ ub{\isacharparenright}\ {\isacharparenleft}Suc\ n{\isacharparenright}\ {\isacharequal}\ {\isacharparenleft}\isanewline
        \ \ \ \ let\ c{\isacharequal}{\isacharparenleft}lb{\isacharplus}ub{\isacharparenright}{\isacharslash}{\isadigit{2}}\isanewline
        \ \ \ \ in\ if\ poly\ p\ lb\ {\isacharasterisk}\ poly\ p\ c\ {\isasymle}\ {\isadigit{0}}\ \isanewline
        \ \ \ \ \ \ \ then\ to{\isacharunderscore}cauchy\ {\isacharparenleft}p{\isacharcomma}\ lb{\isacharcomma}\ c{\isacharparenright}\ n\isanewline
        \ \ \ \ \ \ \ else\ to{\isacharunderscore}cauchy\ {\isacharparenleft}p{\isacharcomma}\ c{\isacharcomma}\ ub{\isacharparenright}\ n{\isacharparenright}{\isachardoublequoteclose}
\end{isabelle}
where \isa{poly p x} evaluates the polynomial \isa{p} at the point \isa{x}.
Note, \isa{rat\ poly\ {\isasymtimes}\ rat\ {\isasymtimes}\ rat} encodes a real
algebraic number here (rather than \isa{int\ poly\ {\isasymtimes}\ float\
  {\isasymtimes}\ float}), as we can embed \isa{int} and \isa{float} into
\isa{rat}.

It can be then shown that the sequence constructed by \isa{to\_cauchy (p, lb,
  ub)} is indeed a Cauchy sequence and the real number represented by this
sequence resides within the interval $[\mathit{lb}, \mathit{ub}]$, provided
$\mathit{lb}<\mathit{ub}$:

\begin{isabelle}
        \isacommand{lemma}\isamarkupfalse%
        \ to{\isacharunderscore}cauchy{\isacharunderscore}cauchy{\isacharcolon}\isanewline
        \ \ \isakeyword{fixes}\ p{\isacharcolon}{\isacharcolon}{\isachardoublequoteopen}rat\ poly{\isachardoublequoteclose}\ \isakeyword{and}\ lb\ ub\ {\isacharcolon}{\isacharcolon}rat\isanewline
        \ \ \isakeyword{assumes}\ {\isachardoublequoteopen}lb{\isacharless}ub{\isachardoublequoteclose}\isanewline
        \ \ \isakeyword{defines}\ {\isachardoublequoteopen}X{\isasymequiv}to{\isacharunderscore}cauchy\ {\isacharparenleft}p{\isacharcomma}lb{\isacharcomma}ub{\isacharparenright}{\isachardoublequoteclose}\isanewline
        \ \ \isakeyword{shows}\ {\isachardoublequoteopen}cauchy\ X{\isachardoublequoteclose}
\end{isabelle}

\begin{isabelle}
        \isacommand{lemma}\isamarkupfalse%
        \ to{\isacharunderscore}cauchy{\isacharunderscore}bound{\isacharcolon}\isanewline
        \ \ \isakeyword{fixes}\ p{\isacharcolon}{\isacharcolon}{\isachardoublequoteopen}rat\ poly{\isachardoublequoteclose}\ \isakeyword{and}\ lb\ ub\ {\isacharcolon}{\isacharcolon}rat\isanewline
        \ \ \isakeyword{defines}\ {\isachardoublequoteopen}X{\isasymequiv}to{\isacharunderscore}cauchy\ {\isacharparenleft}p{\isacharcomma}lb{\isacharcomma}ub{\isacharparenright}{\isachardoublequoteclose}\isanewline
        \ \ \isakeyword{assumes}\ {\isachardoublequoteopen}lb{\isacharless}ub{\isachardoublequoteclose}\ \isanewline
        \ \ \isakeyword{shows}\ {\isachardoublequoteopen}lb\ {\isasymle}\ Real\ X{\isachardoublequoteclose}\ {\isachardoublequoteopen}Real\ X\ {\isasymle}\ ub{\isachardoublequoteclose}
\end{isabelle}
Note, the function \isa{Real} of type \isa{(nat \isasymRightarrow\ rat)
  \isasymRightarrow\ real} constructs a real number from its underlying
representation (i.e. a Cauchy sequence).

Finally, we can finish the definition of \isa{Alg}:
\begin{isabelle}
        \isacommand{definition}\isamarkupfalse%
        \ valid{\isacharunderscore}alg{\isacharcolon}{\isacharcolon}{\isachardoublequoteopen}int\ poly\ {\isasymRightarrow}\ float\ {\isasymRightarrow}\ float\ {\isasymRightarrow}\ bool{\isachardoublequoteclose}\ \isakeyword{where}\isanewline
        \ \ {\isachardoublequoteopen}valid{\isacharunderscore}alg\ p\ lb\ ub\ {\isacharequal}\ {\isacharparenleft}lb\ {\isacharless}\ ub\ {\isasymand}\ poly\ p\ lb\ {\isacharasterisk}\ poly\ p\ ub\ {\isacharless}\ {\isadigit{0}}\ \isanewline
        \ \ \ \ {\isasymand}\ card\ {\isacharparenleft}{\isacharbraceleft}x{\isacharcolon}{\isacharcolon}real{\isachardot}\ poly\ p\ x\ {\isacharequal}\ {\isadigit{0}}\ {\isasymand}\ lb\ {\isacharless}\ x\ {\isasymand}\ x\ {\isacharless}\ ub{\isacharbraceright}{\isacharparenright}\ {\isacharequal}\ {\isadigit{1}}{\isacharparenright}{\isachardoublequoteclose}
\end{isabelle}

\begin{isabelle}
        \isacommand{definition}\isamarkupfalse%
        \ Alg{\isacharcolon}{\isacharcolon}\ {\isachardoublequoteopen}int\ poly\ {\isasymRightarrow}\ float\ {\isasymRightarrow}\ float\ {\isasymRightarrow}\ real{\isachardoublequoteclose}\ \isakeyword{where}\isanewline
        \ \ {\isachardoublequoteopen}Alg\ p\ lb\ ub\ {\isacharequal}\ {\isacharparenleft}if\ valid{\isacharunderscore}alg\ p\ lb\ ub\isanewline
        \ \ \ \ \ \ \ \ then\ Real\ {\isacharparenleft}to{\isacharunderscore}cauchy\ {\isacharparenleft}p{\isacharcomma}\ lb{\isacharcomma}\ ub{\isacharparenright}{\isacharparenright}\isanewline
        \ \ \ \ \ \ \ \ else\ undefined{\isacharparenright}{\isachardoublequoteclose}
\end{isabelle}
where \isa{valid\_alg p lb ub} ensures
\begin{itemize}
        \item \isa{lb < ub},
        \item the polynomial \isa{p} is of different signs (and non-zero) at \isa{lb} and \isa{ub},
        \item the polynomial \isa{p} has exactly one real root within the interval \isa{(lb,ub)}.
\end{itemize}
With the help of \isa{Alg}, we can now encode the real algebraic number $\sqrt{2}$ as
\begin{isabelle}
        Alg [:-2,0,1:] 1 2
\end{isabelle}
where \isa{[:-2,0,1:]} corresponds to the polynomial $-2 x^0+ 0 x^1 + 1 x^2 =
x^2-2$, and $1$ and $2$ are the lower bound and upper bound respectively, such
that $\sqrt{2}$ is the only root of $x^2-2$ within the interval $(1,2)$.

Furthermore, we can formally derive that \isa{Alg\ p\ lb\ ub} is indeed a root
of \isa{p} within the interval \isa{(lb,ub)}:

\begin{isabelle}
        \isacommand{lemma}\isamarkupfalse%
        \ alg{\isacharunderscore}bound{\isacharunderscore}and{\isacharunderscore}root{\isacharcolon}\isanewline
        \ \ \isakeyword{fixes}\ p{\isacharcolon}{\isacharcolon}{\isachardoublequoteopen}int\ poly{\isachardoublequoteclose}\ \isakeyword{and}\ lb\ ub{\isacharcolon}{\isacharcolon}float\isanewline
        \ \ \isakeyword{assumes}\ {\isachardoublequoteopen}valid{\isacharunderscore}alg\ p\ lb\ ub{\isachardoublequoteclose}\isanewline
        \ \ \isakeyword{shows}\ {\isachardoublequoteopen}lb\ {\isacharless}\ Alg\ p\ lb\ ub{\isachardoublequoteclose}\ \isakeyword{and}\ {\isachardoublequoteopen}Alg\ p\ lb\ ub\ {\isacharless}\ ub{\isachardoublequoteclose}\ \isanewline
        \ \ \ \ \isakeyword{and}\ {\isachardoublequoteopen}poly\ {\isacharparenleft}of{\isacharunderscore}int{\isacharunderscore}poly\ p{\isacharparenright}\ {\isacharparenleft}Alg\ p\ lb\ ub{\isacharparenright}\ {\isacharequal}\ {\isadigit{0}}{\isachardoublequoteclose}
\end{isabelle}
where \isa{of\_int\_poly p} embeds the integer polynomial \isa{p} into a real one.

\section{Deciding the Sign of a Univariate Polynomial at Real Algebraic Points}
\label{sec:deciding_sign}
In the previous section, we described how to encode a real algebraic number as
an integer polynomial and two dyadic rational numbers. Now, suppose we have
\[
\sqrt{2} = (x^2-2,1,2)
\]
where $(x^2-2,1,2)$ is abbreviated from \isa{Alg [:-2,0,1:] 1 2} for the sake of
readability. How can we computationally prove that
\[
        P(\sqrt{2}) = 0 \quad \text{where} \quad P(x) = \frac{1}{2} x^2 -1 \ ?
\]
Considering that $\mathbb{R}_{\mathrm{alg}}$ is a computable subfield of $\mathbb{R}$ and has decidable arithmetic and comparison operations, it is natural to evaluate such formulas through algebraic arithmetic:
\[
\begin{split}
&P(\sqrt{2})\\
&= \frac{1}{2}  \times_{\mathrm{alg}}  (x^2-2,1,2)  \times_{\mathrm{alg}} (x^2-2,1,2) -_{\mathrm{alg}} 1\\
&= \frac{1}{2}  \times_{\mathrm{alg}} (x-2,1,3) -_{\mathrm{alg}} 1 \\
&= (x-1,\frac{1}{2},\frac{3}{2}) -_{\mathrm{alg}} 1 \\
&= 0,\\
\end{split}
\]
where $\times_{\mathrm{alg}}$ and $-_{\mathrm{alg}}$ are exact algebraic arithmetic operations that usually involve calculation of bivariate resultants. Although such computations are currently possible in Isabelle/HOL \cite{Algebraic_Numbers_AFP,li2016modular}, they are far from efficient.

In this section, we describe a verified procedure to decide the sign
of univariate polynomials with rational coefficients at real algebraic
points which uses \emph{only} rational (or dyadic rational)
arithmetic rather than costly algebraic arithmetic.

\subsection{The Sturm-Tarski Theorem} \label{sec:sturm_tarski}
We abbreviate $\mathbb{R} \cup \{-\infty,\infty \}$ as $\extR$, the extended real numbers.

\begin{definition}[Tarski Query] \label{def:tarski-query}
        The Tarski query $\TaQ(Q,P,a,b)$ is
        \[ \TaQ(Q,P,a,b) = \sum_{x \in (a,b), P(x)=0} \sgn (Q(x))  \]
        where $a,b \in \extR$, $P, Q \in \mathbb{R}[X]$, $P\neq 0$ and $\sgn : \mathbb{R} \to \{-1,0,1\}$ is the sign function.
\end{definition}

The Sturm-Tarski theorem \cite[Chapter~8]{Mishra}  (or Tarski's theorem \cite[Chapter~2]{real_alg_geo2006}) is essentially an effective way to compute Tarski queries through some remainder sequences:

\begin{theorem}[Sturm-Tarski] \label{thm:sturm-tarski}
        The Sturm-Tarski theorem states
        \[
        \TaQ(Q,P,a,b) = \Var(\SRemS(P,P'Q);a,b)
        \]
        where $P \neq 0$, $P,Q\in \mathbb{R}[X]$, $P'$ is the first
        derivative of $P$, $a,b \in \extR$, $a<b$ and are not roots of
        $P$, $\SRemS(P,P'Q)$ is the signed remainder sequence of $P$
        and $P'Q$, and
        \[
        \begin{split}
        & \Var([p_0, p_1, ..., p_n];a,b) \\
        & = \Var([p_0(a), p_1(a), ..., p_n(a)]) - \Var([p_0(b), p_1(b), ..., p_n(b)]) \\
        \end{split}
        \]
        is the difference in the number of sign variations (after
        removing zeroes) in the polynomial sequence $[p_0 , p_1 , ...,
          p_n ]$ evaluated at $a$ and $b$.
\end{theorem}

Note that the more famous Sturm's theorem, which counts the number of
distinct real roots (of a univariate polynomial) within an interval,
is a special case of the Sturm-Tarski theorem when $Q=1$.

\subsection{A Formal Proof of the Sturm-Tarski Theorem}

Our proof of the Sturm-Tarski theorem in Isabelle is based on Basu et al.\ \cite[Chapter~2]{real_alg_geo2006} and Cohen's formalisation in Coq \cite{cohen_phd}.

The core idea of our formal proof is built around the \emph{Cauchy
  index}.
First defined by Cauchy in 1837, the Cauchy index of a real rational
function encodes deep properties of its roots and poles, and can be
used as the basis of an algebraic method for computing Tarski
queries\footnote{Besides the application described in this section,
  the Cauchy index also plays a critical role in the Routh\textendash
  Hurwitz theorem. Interested readers may
  consult~\cite[Chapter~10,11]{rahman2002analytic} for historical
  notes.}.

\begin{figure}[ht]
        \centering
        \begin{tikzpicture}[scale=1.2,
        declare function={
                f(\x)  = ((\x-4))/((\x-3)*(\x - 1)^2*(\x+1));}
        ]
        %\draw[very thin,color=gray] (-3.1,-4.1) grid (3.9,3.9);
        \draw[->] (-3.2,0) -- (4.2,0) node[right] {$x$};
        %\draw[->] (0,-4.2) -- (0,4.2) node[above] {$y$};
        \draw[domain=-3:-1.1,blue] plot (\x,{f(\x))});
        \draw[domain=-0.88:0.43,blue] plot (\x,{f(\x))});
        \draw[domain=1.47:2.977,blue] plot (\x,{f(\x))});
        \draw[domain=3.02:4,blue] plot (\x,{f(\x))});
        \draw [dashed] (-1,-3) -- (-1,3);
        \draw[fill=black] (-1,0) circle (0.05) node [below right] {$-1$};
%	\node [below right] at (-1,0) {$-1$};
        \draw [dashed] (1,-3) -- (1,3);
        \draw[fill=black] (1,0) circle (0.05) node [below right] {$1$};
%	\node [below right] at (1,0) {$1$};
        \draw [dashed] (3,-3) -- (3,3);
        \draw[fill=black] (3,0) circle (0.05) node [below left] {$3$};
%	\node [below left] at (3,0) {$3$};
        \end{tikzpicture}
        \caption{Graph of the rational function $(x-4) / ((x-3)(x-1)^2(x+1))$} \label{fig:cauchy_index}
\end{figure}
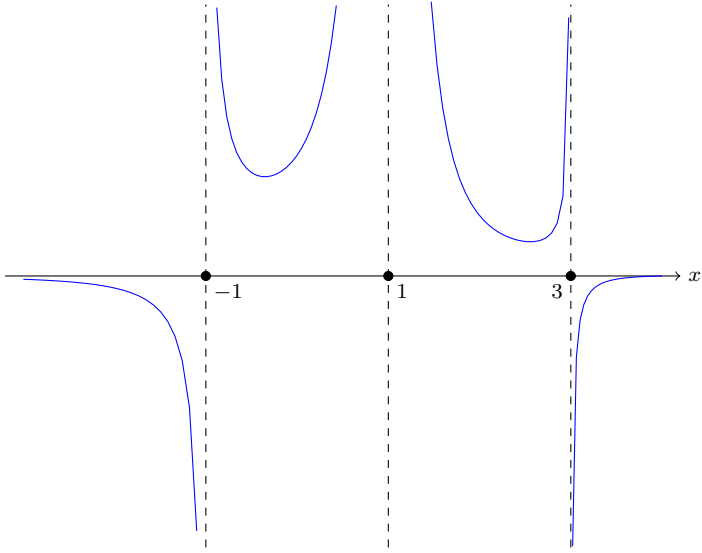

\begin{definition}
        Given $P,Q \in \mathbb{R}[x]$ and $x \in \mathbb{R}$, $\jump(P,Q,x)$ is defined as
        \[
        \jump(P,Q,x) =
        \begin{cases}
        -1 & \mbox{if } \lim_{u \rightarrow x^-} \frac{Q(u)}{P(u)}=\infty \mbox{ and } \lim_{u \rightarrow x^+} \frac{Q(u)}{P(u)}=-\infty\\
        1 & \mbox{if } \lim_{u \rightarrow x^-} \frac{Q(u)}{P(u)}=-\infty \mbox{ and } \lim_{u \rightarrow x^+} \frac{Q(u)}{P(u)}=\infty\\
        0 &  \mbox{ otherwise. }\\
        \end{cases}
        \]
\end{definition}
For example, let $Q=x-4$ and $P=(x-3)(x-1)^2(x+1)$. The graph of $Q/P$ is shown in Fig.~\ref{fig:cauchy_index}. We have
\[
\jump(P,Q,x) =
\begin{cases}
1 & \text{ when } x=-1\\
-1 & \text{ when } x=3\\
0 &  \mbox{ otherwise. }\\
\end{cases}
\]

The Cauchy index \isa{cindex\_poly\ a\ b\ q\ p} is the sum of the jumps of $q/p$ over the interval $(a,b)$:
\begin{isabelle}
        \isacommand{definition}\isamarkupfalse%
        \ cindex{\isacharunderscore}poly{\isacharcolon}{\isacharcolon}\ {\isachardoublequoteopen}real\ {\isasymRightarrow}\ real\ {\isasymRightarrow}\ real\ poly\ {\isasymRightarrow}\ real\ poly\ {\isasymRightarrow}\ int{\isachardoublequoteclose}\ \isanewline
        \ \ \ \ \isakeyword{where}\ \isanewline
        \ \ {\isachardoublequoteopen}cindex{\isacharunderscore}poly\ a\ b\ q\ p{\isasymequiv}\ {\isacharparenleft}{\isasymSum}x{\isasymin}{\isacharbraceleft}x{\isachardot}\ poly\ p\ x{\isacharequal}{\isadigit{0}}\ {\isasymand}\ a\ {\isacharless}\ x\ {\isasymand}\ x\ {\isacharless}\ b{\isacharbraceright}{\isachardot}\ jump{\isacharunderscore}poly\ q\ p\ x{\isacharparenright}{\isachardoublequoteclose}
\end{isabelle}

By case analysis, we can prove a connection between the Tarski query and the Cauchy index:
\begin{isabelle}
        \isacommand{lemma}\isamarkupfalse%
        \ cindex{\isacharunderscore}poly{\isacharunderscore}taq{\isacharcolon}\isanewline
        \ \ \isakeyword{fixes}\ p\ q{\isacharcolon}{\isacharcolon}{\isachardoublequoteopen}real\ poly{\isachardoublequoteclose}\ \isakeyword{and}\ a\ b{\isacharcolon}{\isacharcolon}real\isanewline
        \ \ \isakeyword{shows}\ {\isachardoublequoteopen}taq\ {\isacharbraceleft}x{\isachardot}\ poly\ p\ x\ {\isacharequal}\ {\isadigit{0}}\ {\isasymand}\ a\ {\isacharless}\ x\ {\isasymand}\ x\ {\isacharless}\ b{\isacharbraceright}\ q\isanewline
        \ \ \ \ \ \ \ \ \ \ {\isacharequal}cindex{\isacharunderscore}poly\ a\ b\ {\isacharparenleft}pderiv\ p\ {\isacharasterisk}\ q{\isacharparenright}\ p{\isachardoublequoteclose}
\end{isabelle}
where \isa{taq} is a formal definition of the Tarski query
\begin{isabelle}
        \isacommand{definition}\isamarkupfalse%
        \ taq\ {\isacharcolon}{\isacharcolon}\ {\isachardoublequoteopen}{\isacharprime}a{\isacharcolon}{\isacharcolon}linordered{\isacharunderscore}idom\ set\ {\isasymRightarrow}\ {\isacharprime}a\ poly\ {\isasymRightarrow}\ int{\isachardoublequoteclose}\ \isakeyword{where}\isanewline
        \ \ {\isachardoublequoteopen}taq\ s\ q\ {\isacharequal}\ {\isacharparenleft}{\isasymSum}x{\isasymin}s{\isachardot}\ sign\ {\isacharparenleft}poly\ q\ x{\isacharparenright}{\isacharparenright}{\isachardoublequoteclose}
\end{isabelle}
 and \isa{pderiv p} is the first derivative of \isa{p}.

Moreover, the Cauchy index can be related to Euclidean division
(\isa{mod}) on polynomials by a recurrence:
\begin{isabelle}
        \ cindex{\isacharunderscore}poly{\isacharunderscore}rec{\isacharcolon}\isanewline
        \ \ \isakeyword{fixes}\ p\ q{\isacharcolon}{\isacharcolon}{\isachardoublequoteopen}real\ poly{\isachardoublequoteclose}\ \isakeyword{and}\ a\ b{\isacharcolon}{\isacharcolon}real\isanewline
        \ \ \isakeyword{assumes}\ {\isachardoublequoteopen}a\ {\isacharless}\ b{\isachardoublequoteclose}\ \isakeyword{and}\ {\isachardoublequoteopen}poly\ {\isacharparenleft}p\ {\isacharasterisk}\ q{\isacharparenright}\ a\ {\isasymnoteq}{\isadigit{0}}{\isachardoublequoteclose}\ \isanewline
        \ \ \ \ \ \ \isakeyword{and}\ {\isachardoublequoteopen}poly\ {\isacharparenleft}p\ {\isacharasterisk}\ q{\isacharparenright}\ b\ {\isasymnoteq}{\isadigit{0}}{\isachardoublequoteclose}\isanewline
        \ \ \isakeyword{shows}\ {\isachardoublequoteopen}cindex{\isacharunderscore}poly\ a\ b\ q\ p\ {\isacharequal}\ cross\ {\isacharparenleft}p\ {\isacharasterisk}\ q{\isacharparenright}\ a\ b\ \ \isanewline
        \ \ \ \ \ \ \ \ \ \ {\isacharplus}\ cindex{\isacharunderscore}poly\ a\ b\ {\isacharparenleft}{\isacharminus}\ {\isacharparenleft}p\ mod\ q{\isacharparenright}{\isacharparenright}\ q{\isachardoublequoteclose}
\end{isabelle}
where
\[
\mathrm{cross}\ p\ a\ b =
\begin{cases}
0 & \mbox{if } p(a)p(b) \geq 0\\
1 & \mbox{if } p(a)p(b) < 0 \mbox{ and } p(a)<p(b)\\
-1 & \mbox{if } p(a)p(b) < 0 \mbox{ and } p(a) \geq p(b).\\
\end{cases}
\]
A similar recurrence relation holds for the number of sign variations
of the signed remainder sequences (\isa{changes\_itv\_smods}):
\begin{isabelle}
        \isacommand{lemma}\isamarkupfalse%
        \ changes{\isacharunderscore}itv{\isacharunderscore}smods{\isacharunderscore}rec{\isacharcolon}\isanewline
        \ \ \isakeyword{fixes}\ p\ q{\isacharcolon}{\isacharcolon}{\isachardoublequoteopen}real\ poly{\isachardoublequoteclose}\ \isakeyword{and}\ a\ b{\isacharcolon}{\isacharcolon}real\isanewline
        \ \ \isakeyword{assumes}\ {\isachardoublequoteopen}a\ {\isacharless}\ b{\isachardoublequoteclose}\ \isakeyword{and}\ {\isachardoublequoteopen}poly\ {\isacharparenleft}p\ {\isacharasterisk}\ q{\isacharparenright}\ a\ {\isasymnoteq}\ {\isadigit{0}}{\isachardoublequoteclose}\ \isanewline
        \ \ \ \ \ \ \ \ \isakeyword{and}\ {\isachardoublequoteopen}poly\ {\isacharparenleft}p\ {\isacharasterisk}\ q{\isacharparenright}\ b\ {\isasymnoteq}\ {\isadigit{0}}{\isachardoublequoteclose}\isanewline
        \ \ \isakeyword{shows}\ {\isachardoublequoteopen}changes{\isacharunderscore}itv{\isacharunderscore}smods\ a\ b\ p\ q\ {\isacharequal}\ cross\ {\isacharparenleft}p\ {\isacharasterisk}\ q{\isacharparenright}\ a\ b\ \isanewline
        \ \ \ \ \ \ \ \ \ \ {\isacharplus}\ changes{\isacharunderscore}itv{\isacharunderscore}smods\ a\ b\ q\ {\isacharparenleft}{\isacharminus}\ {\isacharparenleft}p\ mod\ q{\isacharparenright}{\isacharparenright}{\isachardoublequoteclose}
\end{isabelle}
where \isa{changes\_itv\_smods} is defined as
\begin{isabelle}
        \isacommand{definition}\isamarkupfalse%
        \ changes{\isacharunderscore}itv{\isacharunderscore}smods{\isacharcolon}{\isacharcolon}\ \isanewline
        \ \ \ \ {\isachardoublequoteopen}real\ {\isasymRightarrow}\ real\ {\isasymRightarrow}\ real\ poly\ {\isasymRightarrow}\ real\ poly\ {\isasymRightarrow}\ int{\isachardoublequoteclose}\ \isakeyword{where}\isanewline
        \ \
        {\isachardoublequoteopen}changes{\isacharunderscore}itv{\isacharunderscore}smods\
        a\ b\ p\ q\ {\isacharequal}\ {\isacharparenleft}\isanewline
        \ \ \ \ \ \ let\ \isanewline
        \ \ \ \ \ \ \ \ ps\ {\isacharequal}\ smods\ p\ q\ \isanewline
        \ \ \ \ \ \ in\ \isanewline
        \ \ \ \ \ \ \ \ changes{\isacharunderscore}poly{\isacharunderscore}at\ ps\ a\ {\isacharminus}\ changes{\isacharunderscore}poly{\isacharunderscore}at\ ps\ b{\isacharparenright}{\isachardoublequoteclose}
\end{isabelle}
and the signed remainder sequence (\isa{smods}) is defined as
\begin{isabelle}
        \isacommand{function}\isamarkupfalse%
        \ smods{\isacharcolon}{\isacharcolon}\ {\isachardoublequoteopen}real\ poly\ {\isasymRightarrow}\ real\ poly\ {\isasymRightarrow}\ {\isacharparenleft}real\ poly{\isacharparenright}\ list{\isachardoublequoteclose}\ \isakeyword{where}\isanewline
        \ \ {\isachardoublequoteopen}smods\ p\ q{\isacharequal}\ {\isacharparenleft}if\ p\ {\isacharequal}\ {\isadigit{0}}\ then\ \isanewline
        \ \ \ \ \ \ \ \ \ \ \ \ \ \ \ \ \ {\isacharbrackleft}{\isacharbrackright}\ \isanewline
        \ \ \ \ \ \ \ \ \ \ \ \ \ \ \ else\ \isanewline
        \ \ \ \ \ \ \ \ \ \ \ \ \ \ \ \ \ p\ {\isacharhash}\ {\isacharparenleft}smods\ q\ {\isacharparenleft}{\isacharminus}{\isacharparenleft}p\ mod\ q{\isacharparenright}{\isacharparenright}{\isacharparenright}{\isacharparenright}{\isachardoublequoteclose}
\end{isabelle}
and \isa{changes\_poly\_at ps a} returns the number of sign changes when evaluating a list of polynomials (\isa{ps}) at \isa{a}.

Finally, by combining \isa{cindex\_poly\_taq}, \isa{cindex\_poly\_rec} and \isa{changes\_itv\_smods\_rec}, we derive the Sturm-Tarski theorem:
\begin{isabelle}
        \isacommand{theorem}\isamarkupfalse%
        \ sturm{\isacharunderscore}tarski{\isacharunderscore}interval{\isacharcolon}\ \isanewline
        \ \ \isakeyword{fixes}\ p\ q{\isacharcolon}{\isacharcolon}{\isachardoublequoteopen}real\ poly{\isachardoublequoteclose}\ \isakeyword{and}\ a\ b{\isacharcolon}{\isacharcolon}real\isanewline
        \ \ \isakeyword{assumes}\ {\isachardoublequoteopen}a\ {\isacharless}\ b{\isachardoublequoteclose}\ \isakeyword{and}\ {\isachardoublequoteopen}poly\ p\ a\ {\isasymnoteq}\ {\isadigit{0}}{\isachardoublequoteclose}\ \isakeyword{and}\ {\isachardoublequoteopen}poly\ p\ b\ {\isasymnoteq}\ {\isadigit{0}}{\isachardoublequoteclose}\isanewline
        \ \ \isakeyword{shows}\ {\isachardoublequoteopen}taq\ {\isacharbraceleft}x{\isachardot}\ poly\ p\ x\ {\isacharequal}\ {\isadigit{0}}\ {\isasymand}\ a\ {\isacharless}\ x\ {\isasymand}\ x\ {\isacharless}\ b{\isacharbraceright}\ q\ \isanewline
        \ \ \ \ \ \ {\isacharequal}\ changes{\isacharunderscore}itv{\isacharunderscore}smods\ a\ b\ p\ {\isacharparenleft}pderiv\ p\ {\isacharasterisk}\ q{\isacharparenright}{\isachardoublequoteclose}
\end{isabelle}

Note, this is just the bounded case of the Sturm-Tarski theorem. Proofs for the
unbounded and half-bounded cases are similar.

\subsection{Sign Determination through the Sturm-Tarski Theorem}
\label{sec:deciding_sign_sturm_tarski}

Given a polynomial $q$ with rational coefficients and our encoding of a real algebraic number $\alpha$
\[
\alpha = (p, \mathit{lb}, \mathit{ub})
\]
where $p$ is an integer polynomial, and $\mathit{lb}$ and $\mathit{ub}$ are
dyadic rationals, we can effectively decide the sign of $q(\alpha)$ using the
Sturm-Tarski theorem, provided \isa{valid\_alg p lb ub} holds. The rationale
behind is that $\isa{valid\_alg p lb ub}$ ensures $\alpha$ is the only root of
$p$ within the interval $(\mathit{lb}, \mathit{ub})$, hence
\begin{align*}
\sgn (q(\alpha)) &= \sum_{x \in (\mathit{lb},\mathit{ub}), p(x)=0} \sgn (q(x))\\
&= \TaQ(q,p,lb,ub)\\
&= \Var(\SRemS(p,p'q);\mathit{lb},\mathit{ub}).
\end{align*}
Importantly, it can be observed that evaluating
$\Var(\SRemS(p,p'q);\mathit{lb},\mathit{ub})$ requires only rational
arithmetic rather than costly algebraic arithmetic.

To be even more efficient, we refine the procedure further to make use of dyadic
rational arithmetic. The main advantage of dyadic rational arithmetic over
rational arithmetic are reduced normalization steps and possible bit-level
operations. For example, consider two rational numbers $\frac{a_1}{b_1}$ and
$\frac{a_1}{b_2}$ where $a_1,b_1,a_2,b_2 \in \mathbb{Z}$, their sum is
\[
\begin{split}
        \frac{a_1}{b_1} + \frac{a_2}{b_2} = \frac{a_1 b_2 + a_2 b_1}{b_1 b_2} = \frac{(a_1 b_2 + a_2 b_1)/c}{(b_1 b_2)/c} \quad \\
        \text{where} \ c=\gcd(a_1 b_2 + a_2 b_1,b_1 b_2).
\end{split}
\]
To counter the growth in the size of representations, we usually need
to normalize the result by factoring out the gcd. Such gcd operations
can be the source of major computational expense. Thankfully, they are
unnecessary in the context of dyadic rationals. The sum of two dyadic
rationals $(a_1,e_1)$ and $(a_2,e_2)$ where $a_1,e_1,a_2,b_2
\in \mathbb{Z}$ is
\[
a_1  2^{e_1} + a_2  2^{e_2} =
\begin{cases}
(a_1 2^{e_1-e_2}+a_2) 2^{e_2} & \mbox{if } e_1>e_2\\
(a_1+a_2 2^{e_2-e_1}) 2^{e_1}  & \mbox{otherwise.} \\
\end{cases}
\]
Moreover, multiplications by powers of two, such as $a_1 2^{e_1-e_2}$, can be
optimised by shift operations.

However, the problem with dyadic rational numbers is that they do not have the
division operation (e.g. $1 \times 2^0$ divided by $3 \times 2^0$ is no longer
a dyadic rational), hence they do not form a field, while Euclidean division only
works for polynomials over a field. This problem can be solved if we switch from
Euclidean division (\isa{mod} and \isa{div}):
\begin{multline*}
        P = (P \div Q)\, Q + (P \mod Q) \  \text{ and } \  (Q = 0  \vee \deg(P \mod Q) < \deg(Q))
\end{multline*}
to pseudo-division (\isa{pmod} and \isa{pdiv})~\cite{de2013computation}:
\begin{multline*}
        \mathrm{lc}(Q)^{1+\deg(P) -\deg(Q)} P = (P \pdiv Q)\, Q + (P \pmod Q) \\
        \text{ and } \  (Q = 0  \vee \deg(P \mod Q) < \deg(Q)) \\
        \text{where lc$(Q)$ is the leading coefficient of $Q$,}
\end{multline*}
since pseudo-division can be carried out by polynomials over an integral domain (rather than a field).

Based on pseudo-division, the signed pseudo-remainder sequence ($\SPRemS$) can be defined:
\begin{isabelle}
        \isacommand{function}\isamarkupfalse%
        \ spmods\ {\isacharcolon}{\isacharcolon}\ {\isachardoublequoteopen}{\isacharprime}a{\isacharcolon}{\isacharcolon}idom\ poly\ {\isasymRightarrow}\ {\isacharprime}a\ poly\ {\isasymRightarrow}\ {\isacharparenleft}{\isacharprime}a\ poly{\isacharparenright}\ list{\isachardoublequoteclose}\ \isakeyword{where}\isanewline
        \ \ {\isachardoublequoteopen}spmods\ p\ q{\isacharequal}\ {\isacharparenleft}if\ p{\isacharequal}{\isadigit{0}}\ then\ {\isacharbrackleft}{\isacharbrackright}\ else\ \isanewline
        \ \ \ \ \ \ let\ \isanewline
        \ \ \ \ \ \ \ \ m{\isacharequal}{\isacharparenleft}if\ even{\isacharparenleft}degree\ p{\isacharplus}{\isadigit{1}}{\isacharminus}degree\ q{\isacharparenright}\ then\ {\isacharminus}{\isadigit{1}}\ else\ {\isacharminus}lead{\isacharunderscore}coeff\ q{\isacharparenright}\ \isanewline
        \ \ \ \ \ \ in\ \ \ \ \ \isanewline
        \ \ \ \ \ \ \ \ Cons\ p\ {\isacharparenleft}spmods\ q\ {\isacharparenleft}smult\ m\ {\isacharparenleft}p\ pmod\ q{\isacharparenright}{\isacharparenright}{\isacharparenright}{\isacharparenright}{\isachardoublequoteclose}
\end{isabelle}
where \isa{smult} is the scalar product on polynomials and \isa{lead\_coeff q} is the leading coefficient of \isa{q}. Accordingly, the function to count the difference in sign variations can be refined:
\begin{isabelle}
        \isacommand{definition}\isamarkupfalse%
        \ changes{\isacharunderscore}itv{\isacharunderscore}spmods{\isacharcolon}{\isacharcolon}\ \isanewline
        \ \ {\isachardoublequoteopen}{\isacharprime}a\ {\isacharcolon}{\isacharcolon}linordered{\isacharunderscore}idom\ {\isasymRightarrow}\ {\isacharprime}a\ {\isasymRightarrow}\ {\isacharprime}a\ poly\ {\isasymRightarrow}\ {\isacharprime}a\ poly\ {\isasymRightarrow}\ int{\isachardoublequoteclose}\ \isakeyword{where}\isanewline
        \ \ {\isachardoublequoteopen}changes{\isacharunderscore}itv{\isacharunderscore}spmods\ a\ b\ p\ q{\isacharequal}\ {\isacharparenleft}let\ ps\ {\isacharequal}\ spmods\ p\ q\ in\ \isanewline
        \ \ \ \ \ \ \ \ changes{\isacharunderscore}poly{\isacharunderscore}at\ ps\ a\ {\isacharminus}\ changes{\isacharunderscore}poly{\isacharunderscore}at\ ps\ b{\isacharparenright}{\isachardoublequoteclose}
\end{isabelle}
and linked to the previous one based on signed remainder sequences ($\SRemS$):
\begin{isabelle}
        \isacommand{lemma}\isamarkupfalse%
        \ changes{\isacharunderscore}spmods{\isacharunderscore}smods{\isacharcolon}\isanewline
        \ \ \isakeyword{fixes}\ p\ q{\isacharcolon}{\isacharcolon}{\isachardoublequoteopen}float\ poly{\isachardoublequoteclose}\ \isakeyword{and}\ a\ b{\isacharcolon}{\isacharcolon}{\isachardoublequoteopen}float{\isachardoublequoteclose}\isanewline
        \ \ \isakeyword{shows}\ {\isachardoublequoteopen}changes{\isacharunderscore}itv{\isacharunderscore}spmods\ a\ b\ p\ q\ \isanewline
        \ \ \ \ \ \ \ \ \ \ {\isacharequal}\ changes{\isacharunderscore}itv{\isacharunderscore}smods\ {\isacharparenleft}real{\isacharunderscore}of{\isacharunderscore}float\ a{\isacharparenright}\ {\isacharparenleft}real{\isacharunderscore}of{\isacharunderscore}float\ b{\isacharparenright}\ \isanewline
        \ \ \ \ \ \ \ \ \ \ \ \ \ \ \ \ \ \ \ \ \ \ \ \ \ \ \ \ \ \ {\isacharparenleft}of{\isacharunderscore}float{\isacharunderscore}poly\ p{\isacharparenright}\ {\isacharparenleft}of{\isacharunderscore}float{\isacharunderscore}poly\ q{\isacharparenright}{\isachardoublequoteclose}
\end{isabelle}
where \isa{real\_of\_float} embeds a \isa{float} into \isa{real} and \isa{of\_float\_poly} coverts a \isa{float poly} (i.e. polynomial with dyadic rational coefficients) to a \isa{real poly} by embedding each of the coefficients into \isa{real}.

Finally, we define a function \isa{sgn\_at} that returns the sign of a univariate polynomial at some point:
\begin{isabelle}
        \isacommand{definition}\isamarkupfalse%
        \ {\isachardoublequoteopen}{\isacharparenleft}sgn{\isacharunderscore}at{\isacharcolon}{\isacharcolon}real\ poly{\isasymRightarrow}real{\isasymRightarrow}real{\isacharparenright}\ {\isacharequal}\ {\isacharparenleft}{\isasymlambda}q\ x{\isachardot}\ sgn\ {\isacharparenleft}poly\ q\ x{\isacharparenright}{\isacharparenright}{\isachardoublequoteclose}
\end{isabelle}
Note, for now, if either \isa{x} or any coefficient of \isa{q} is an
irrational real number (e.g. an irrational real algebraic number),
evaluating \isa{sgn\_at q x} will raise an exception, as Isabelle/HOL,
by default, only supports rational arithmetic. Although we can
eliminate some such exceptions by loading any of the recent algebraic
arithmetic libraries \cite{Algebraic_Numbers_AFP,li2016modular}, we
consider exact algebraic arithmetic too slow for our purpose as stated
at the beginning of Sec.~\ref{sec:deciding_sign}. Alternatively, by
proving some code equations, we can restore the executability of
\isa{sgn\_at q x} when \isa{x} is constructed by \isa{Alg p lb ub} and
coefficients of \isa{q} are rational reals:
\begin{isabelle}
        \isacommand{lemma}\isamarkupfalse%
        \ sgn{\isacharunderscore}at{\isacharunderscore}code{\isacharunderscore}alg{\isacharbrackleft}code{\isacharbrackright}{\isacharcolon}\ \isanewline
        \ \ \isakeyword{fixes}\ q{\isacharcolon}{\isacharcolon}{\isachardoublequoteopen}real\ poly{\isachardoublequoteclose}\ \isakeyword{and}\ p{\isacharcolon}{\isacharcolon}{\isachardoublequoteopen}int\ poly{\isachardoublequoteclose}\ \isakeyword{and}\ lb\ ub{\isacharcolon}{\isacharcolon}float\isanewline
        \ \ \isakeyword{shows}\ {\isachardoublequoteopen}sgn{\isacharunderscore}at\ q\ {\isacharparenleft}Alg\ p\ lb\ ub{\isacharparenright}\ {\isacharequal}\ {\isacharparenleft}\ \isanewline
        \ \ \ \ if\ valid{\isacharunderscore}alg\ p\ lb\ ub\ {\isasymand}\ {\isacharparenleft}{\isasymforall}x{\isasymin}set\ {\isacharparenleft}coeffs\ q{\isacharparenright}{\isachardot}\ is{\isacharunderscore}rat\ x{\isacharparenright}\ then\ \isanewline
        \ \ \ \ \ \ {\isacharparenleft}let\isanewline
        \ \ \ \ \ \ \ \ \ \ p{\isacharprime}{\isacharcolon}{\isacharcolon}float\ poly{\isacharequal}of{\isacharunderscore}int{\isacharunderscore}poly\ p{\isacharsemicolon}\isanewline
        \ \ \ \ \ \ \ \ \ \ q{\isacharprime}{\isacharcolon}{\isacharcolon}float\ poly{\isacharequal}of{\isacharunderscore}int{\isacharunderscore}poly\ {\isacharparenleft}int{\isacharunderscore}poly\ q{\isacharparenright}\isanewline
        \ \ \ \ \ \ \ in\isanewline
        \ \ \ \ \ \ of{\isacharunderscore}int\ {\isacharparenleft}changes{\isacharunderscore}itv{\isacharunderscore}spmods\ lb\ ub\ p{\isacharprime}\ {\isacharparenleft}pderiv\ p{\isacharprime}\ {\isacharasterisk}\ q{\isacharprime}{\isacharparenright}{\isacharparenright}{\isacharparenright}\isanewline
        \ \ \ \ else\ Code{\isachardot}abort\ {\isacharparenleft}STR\ {\isacharprime}{\isacharprime}Invalid\ sgn{\isacharunderscore}at{\isacharprime}{\isacharprime}{\isacharparenright}\ \isanewline
        \ \ \ \ \ \ \ \ {\isacharparenleft}{\isasymlambda}{\isacharunderscore}{\isachardot}\ sgn{\isacharunderscore}at\ q\ {\isacharparenleft}Alg\ p\ lb\ ub{\isacharparenright}{\isacharparenright}{\isacharparenright}{\isachardoublequoteclose}
\end{isabelle}
where
\begin{itemize}
        \item
          \isa{{\isasymforall}x{\isasymin}set\ {\isacharparenleft}coeffs\ q{\isacharparenright}{\isachardot}\ is{\isacharunderscore}rat\ x}
          checks if each coefficient of \isa{q} is rational,
        \item \isa{of\_int\_poly} converts an integer polynomial into
          a dyadic rational one,
        \item \isa{int\_poly} clears denominators in the coefficients
          by multiplying each coefficient by the least common multiple
          (of the denominators),
        \item \isa{Code.abort} throws an exception, if either
          $(p,\mathit{lb},\mathit{ub})$ is an invalid representation
          of a real algebraic number or the polynomial \isa{q} has any
          non-rational coefficient.
\end{itemize}
And note that evaluating \isa{changes{\isacharunderscore}itv{\isacharunderscore}spmods\ lb\ ub\ p{\isacharprime}\ {\isacharparenleft}pderiv\ p{\isacharprime}\ {\isacharasterisk}\ q{\isacharprime}{\isacharparenright}} requires only dyadic arithmetic, which is much more efficient than exact algebraic arithmetic.

Moreover, the executability of \isa{valid\_alg} is restored similarly as well:
\begin{isabelle}
        \isacommand{lemma}\isamarkupfalse%
        \ {\isacharbrackleft}code{\isacharbrackright}{\isacharcolon}\isanewline
        \ \ \isakeyword{fixes}\ p{\isacharcolon}{\isacharcolon}{\isachardoublequoteopen}int\ poly{\isachardoublequoteclose}\ \isakeyword{and}\ lb\ ub{\isacharcolon}{\isacharcolon}float\ \isanewline
        \ \ \isakeyword{shows}\ {\isachardoublequoteopen}valid{\isacharunderscore}alg\ p\ lb\ ub\ {\isacharequal}\ {\isacharparenleft}lb\ {\isacharless}\ ub\ \isanewline
        \ \ \ \ {\isasymand}\ {\isacharparenleft}sgn\ {\isacharparenleft}poly\ {\isacharparenleft}of{\isacharunderscore}int{\isacharunderscore}poly\ p{\isacharparenright}\ lb{\isacharparenright}\ {\isacharasterisk}\ sgn\ {\isacharparenleft}poly\ {\isacharparenleft}of{\isacharunderscore}int{\isacharunderscore}poly\ p{\isacharparenright}\ ub{\isacharparenright}\ {\isacharless}\ {\isadigit{0}}{\isacharparenright}\ \isanewline
        \ \ \ \ {\isasymand}\ changes{\isacharunderscore}itv{\isacharunderscore}spmods\ lb\ ub\ {\isacharparenleft}of{\isacharunderscore}int{\isacharunderscore}poly\ p{\isacharparenright}\ {\isacharparenleft}pderiv\ {\isacharparenleft}of{\isacharunderscore}int{\isacharunderscore}poly\ p{\isacharparenright}{\isacharparenright}\ {\isacharequal}\ {\isadigit{1}}{\isacharparenright}{\isachardoublequoteclose}
\end{isabelle}
where
\begin{isabelle}
        changes{\isacharunderscore}itv{\isacharunderscore}spmods\ lb\ ub\ {\isacharparenleft}of{\isacharunderscore}int{\isacharunderscore}poly\ p{\isacharparenright}\ {\isacharparenleft}pderiv\ {\isacharparenleft}of{\isacharunderscore}int{\isacharunderscore}poly\ p{\isacharparenright}{\isacharparenright}\ {\isacharequal}\ {\isadigit{1}}
\end{isabelle}
checks if the polynomial $p$ has exactly one real root within the
interval $(\mathit{lb},\mathit{ub})$ by exploiting Sturm's theorem
(a special case of our formalised Sturm-Tarski theorem).

After restoring executability of \isa{sgn\_at} on real algebraic numbers, we can now check the sign of $P(x)=\frac{1}{2}x^2-1$ at $\sqrt{2}$ by typing the following command:
\begin{isabelle}
        \isacommand{value}\isamarkupfalse%
        \ {\isachardoublequoteopen}sgn{\isacharunderscore}at\ {\isacharbrackleft}{\isacharcolon}{\isacharminus}{\isadigit{1}}{\isacharcomma}{\isadigit{0}}{\isacharcomma}{\isadigit{1}}{\isacharslash}{\isadigit{2}}{\isacharcolon}{\isacharbrackright}\ {\isacharparenleft}Alg\ {\isacharbrackleft}{\isacharcolon}{\isacharminus}{\isadigit{2}}{\isacharcomma}{\isadigit{0}}{\isacharcomma}{\isadigit{1}}{\isacharcolon}{\isacharbrackright}\ {\isadigit{1}}\ {\isadigit{2}}{\isacharparenright}{\isachardoublequoteclose}
\end{isabelle}
which returns 0 (i.e. $P(\sqrt{2})=0$).

\subsection{Remark}

A formal proof of the Sturm-Tarski theorem is not new among proof assistants:
it has been formalised in PVS \cite{narkawicz2015formally} and Coq
\cite{cohen_phd}. However, as far as we know, we are the first to exploit this
theorem to build a verified sign determination procedure of real algebraic
numbers, which uses only rational or dyadic rational arithmetic.

Real algebraic numbers are essential in symbolic computing, and well studied. In
general, exact real algebraic arithmetic is rarely used in modern computer
algebra systems due to its extreme inefficiency. For example, consider the
problem of isolating the real roots of a polynomial with real algebraic
coefficients. Modern approaches usually use sophisticated techniques to soundly
approximate those coefficients to a certain precision rather than carrying out
exact algebraic arithmetic
\cite{sagraloff2010general,cheng2007complete,strzebonski2006cylindrical},
relying on exact symbolic procedures as a fall-back in degenerate cases.

Following these efficient modern approaches, our sign determination procedure
can be improved in at least the following ways:
\begin{itemize}
\item Sophisticated interval arithmetic can be used to decide the sign before
  resorting to a remainder sequence, as has been done in Z3
  \cite{de2013computation}. This approach should help when the sign is non-zero.
\item Pseudo-division, which we are currently using for building remainder
  sequences, is not good for controlling coefficients growth. More sophisticated
  approaches, such as subresultant sequences and modular methods, can be used to
  optimise the calculation of remainder sequences.
\end{itemize}

\section{The Formal Development of the Decision Procedure} \label{sec:main_proof}
In this section, we describe the main proof underlying our tactic.

\subsection{Parsing Formulas}

The first step of our tactic is to parse the target formula into a
structured form. This process is usually referred as reification
\cite{chaieb2008automated} in Isabelle/HOL. More specifically, given an
Isabelle/HOL term $e$ of type $\tau$, we define a (more structured) datatype
$\delta$ and an interpretation function $\mathit{interp}$ of type $\delta
\Rightarrow \tau \ \mathit{list} \Rightarrow \tau$, such that for some $e`$ of
type $\delta$
\[
        e = \mathit{interp}\ e`\ \mathit{xs}
\]
where $\mathit{xs}$ is a list of free variables in $e$. Subsequently, instead of
directly dealing with $e$, we now convert it into a more pleasant form
$\mathit{interp}\ e`\ \mathit{xs}$ where $e`$ is in fact a formal language that
captures the structure of $e$.

The datatypes we defined to capture the structure of target univariate formulas
are as follows:
\begin{isabelle}
        \isacommand{datatype}\isamarkupfalse%
        \ num\ {\isacharequal}\ C\ real\ %
        \isamarkupcmt{Constant%
        }
        \isanewline
        \ \ \ \ {\isacharbar}\ Var\ nat\ %
        \isamarkupcmt{Variable index%
        }
        \isanewline
        \ \ \ \ {\isacharbar}\ Add\ num\ num\ {\isacharbar}\ Minus\ num\ {\isacharbar}\ Mul\ num\ num\ {\isacharbar}\ Power\ num\ nat
\end{isabelle}
\begin{isabelle}
        \isacommand{datatype}\isamarkupfalse%
        \ norm{\isacharunderscore}num{\isadigit{2}}\ {\isacharequal}\ \isanewline
        \ \ \ \ Pol\ {\isachardoublequoteopen}int\ poly{\isachardoublequoteclose}\ nat\ %
        \isamarkupcmt{an integer polynomial and its variable index%
        }
        \isanewline
        \ \ \ \ {\isacharbar}\ Const\ real\ %
        \isamarkupcmt{constant%
        }
        \isanewline
        \ \ \ \ {\isacharbar}\ Abnorm\ num\ %
        \isamarkupcmt{in case of anomalies (e.g., bivariate)%
        }
\end{isabelle}
 \begin{isabelle}
        \isacommand{datatype}\isamarkupfalse%
        \ qf{\isacharunderscore}form{\isadigit{2}}\ {\isacharequal}\ \ \isanewline
        \ \ \ \ Pos\ norm{\isacharunderscore}num{\isadigit{2}}\ %
        \isamarkupcmt{is positive%
        }
        {\isacharbar}\ Zero\ norm{\isacharunderscore}num{\isadigit{2}}\ %
        \isamarkupcmt{is zero%
        }
        \ \isanewline
        \ \ \ \ {\isacharbar}\ Neg\ qf{\isacharunderscore}form{\isadigit{2}}\ %
        \isamarkupcmt{negation%
        }
        \ \isanewline
        \ \ \ \ {\isacharbar}\ Conj\ qf{\isacharunderscore}form{\isadigit{2}}\ qf{\isacharunderscore}form{\isadigit{2}}\ %
        \isamarkupcmt{conjunction%
        }
        \ \isanewline
        \ \ \ \ {\isacharbar}\ Disj\ qf{\isacharunderscore}form{\isadigit{2}}\ qf{\isacharunderscore}form{\isadigit{2}}\ %
        \isamarkupcmt{disjunction%
        }
        \ \isanewline
        \ \ \ \ {\isacharbar}\ T\ %
        \isamarkupcmt{true%
        }
        \ {\isacharbar}\ F\ %
        \isamarkupcmt{false%
        }
\end{isabelle}
\begin{isabelle}
        \isacommand{datatype}\isamarkupfalse%
        \ norm{\isacharunderscore}form{\isadigit{2}}\ {\isacharequal}\ \isanewline
        \ \ \ \ QF\ qf{\isacharunderscore}form{\isadigit{2}}\ %
        \isamarkupcmt{quantifier free%
        }
        \isanewline
        \ \ \ \ {\isacharbar}\ ExQ\ norm{\isacharunderscore}form{\isadigit{2}}\ %
        \isamarkupcmt{existential%
        }
        \ \isanewline
        \ \ \ \ {\isacharbar}\ AllQ\ norm{\isacharunderscore}form{\isadigit{2}}\ %
        \isamarkupcmt{universal%
        }
\end{isabelle}
and the interpretation functions:
\begin{isabelle}
        \isacommand{fun}\isamarkupfalse%
        \ num{\isacharunderscore}interp{\isacharcolon}{\isacharcolon}\ {\isachardoublequoteopen}num\ {\isasymRightarrow}\ real\ list\ {\isasymRightarrow}\ real{\isachardoublequoteclose}\
        \isakeyword{where}\isanewline
        \ \ {\isachardoublequoteopen}num{\isacharunderscore}interp\ {\isacharparenleft}C\ i{\isacharparenright}\ vs\ {\isacharequal}\ i{\isachardoublequoteclose}{\isacharbar}\ \isanewline
        \ \ {\isachardoublequoteopen}num{\isacharunderscore}interp\ {\isacharparenleft}Var\ v{\isacharparenright}\ vs\ {\isacharequal}\ vs{\isacharbang}v{\isachardoublequoteclose}{\isacharbar}\ \isanewline
        \ \ {\isachardoublequoteopen}num{\isacharunderscore}interp\ {\isacharparenleft}Add\ num{\isadigit{1}}\ num{\isadigit{2}}{\isacharparenright}\ vs\ {\isacharequal}\ num{\isacharunderscore}interp\ num{\isadigit{1}}\ vs\ {\isacharplus}\ num{\isacharunderscore}interp\ num{\isadigit{2}}\ vs\ {\isachardoublequoteclose}{\isacharbar}\isanewline
        \ \ {\isachardoublequoteopen}num{\isacharunderscore}interp\ {\isacharparenleft}Minus\ num{\isacharparenright}\ vs\ {\isacharequal}\ {\isacharminus}\ num{\isacharunderscore}interp\ num\ vs\ {\isachardoublequoteclose}{\isacharbar}\ \isanewline
        \ \ {\isachardoublequoteopen}num{\isacharunderscore}interp\ {\isacharparenleft}Mul\ num{\isadigit{1}}\ num{\isadigit{2}}{\isacharparenright}\ vs\ {\isacharequal}\ num{\isacharunderscore}interp\ num{\isadigit{1}}\ vs\ {\isacharasterisk}\ num{\isacharunderscore}interp\ num{\isadigit{2}}\ vs\ {\isachardoublequoteclose}{\isacharbar}\ \ \isanewline
        \ \ {\isachardoublequoteopen}num{\isacharunderscore}interp\ {\isacharparenleft}Power\ num\ n{\isacharparenright}\ vs\ {\isacharequal}\ {\isacharparenleft}num{\isacharunderscore}interp\ num\ vs{\isacharparenright}{\isacharcircum}n{\isachardoublequoteclose}
  \end{isabelle}
 \begin{isabelle}
        \isacommand{fun}\isamarkupfalse%
        \ norm{\isacharunderscore}num{\isadigit{2}}{\isacharunderscore}interp\ {\isacharcolon}{\isacharcolon}\ {\isachardoublequoteopen}norm{\isacharunderscore}num{\isadigit{2}}\ {\isasymRightarrow}\ real\ list\ {\isasymRightarrow}\ real{\isachardoublequoteclose}\ \isakeyword{where}\isanewline
        \ \ {\isachardoublequoteopen}norm{\isacharunderscore}num{\isadigit{2}}{\isacharunderscore}interp\ {\isacharparenleft}Pol\ p\ v{\isacharparenright}\ vs\ {\isacharequal}\ poly\ {\isacharparenleft}of{\isacharunderscore}int{\isacharunderscore}poly\ p{\isacharparenright}\ {\isacharparenleft}vs{\isacharbang}v{\isacharparenright}{\isachardoublequoteclose}{\isacharbar}\isanewline
        \ \ {\isachardoublequoteopen}norm{\isacharunderscore}num{\isadigit{2}}{\isacharunderscore}interp\ {\isacharparenleft}Const\ c{\isacharparenright}\ vs\ {\isacharequal}\ c{\isachardoublequoteclose}{\isacharbar}\isanewline
        \ \ {\isachardoublequoteopen}norm{\isacharunderscore}num{\isadigit{2}}{\isacharunderscore}interp\ {\isacharparenleft}Abnorm\ num{\isacharparenright}\ vs\ {\isacharequal}\ num{\isacharunderscore}interp\ num\ vs{\isachardoublequoteclose}\ %
        \isamarkupcmt{anomaly%
        }
\end{isabelle}

\begin{isabelle}
        \isacommand{fun}\isamarkupfalse%
        \ qf{\isacharunderscore}form{\isadigit{2}}{\isacharunderscore}interp{\isacharcolon}{\isacharcolon}\ {\isachardoublequoteopen}qf{\isacharunderscore}form{\isadigit{2}}\ {\isasymRightarrow}\ real\ list\ {\isasymRightarrow}\ bool{\isachardoublequoteclose}\ \isakeyword{where}\ \isanewline
        \ \ {\isachardoublequoteopen}qf{\isacharunderscore}form{\isadigit{2}}{\isacharunderscore}interp\ {\isacharparenleft}Pos\ norm{\isacharunderscore}num{\isacharparenright}\ vs\ {\isacharequal}\ {\isacharparenleft}norm{\isacharunderscore}num{\isadigit{2}}{\isacharunderscore}interp\ norm{\isacharunderscore}num\ vs\ {\isachargreater}\ {\isadigit{0}}{\isacharparenright}{\isachardoublequoteclose}{\isacharbar}\isanewline
        \ \ {\isachardoublequoteopen}qf{\isacharunderscore}form{\isadigit{2}}{\isacharunderscore}interp\ {\isacharparenleft}Zero\ norm{\isacharunderscore}num{\isacharparenright}\ vs\ {\isacharequal}\ {\isacharparenleft}norm{\isacharunderscore}num{\isadigit{2}}{\isacharunderscore}interp\ norm{\isacharunderscore}num\ vs\ {\isacharequal}\ {\isadigit{0}}{\isacharparenright}{\isachardoublequoteclose}{\isacharbar}\isanewline
        \ \ {\isachardoublequoteopen}qf{\isacharunderscore}form{\isadigit{2}}{\isacharunderscore}interp\ {\isacharparenleft}Neg\ qf{\isacharunderscore}form{\isacharparenright}\ vs\ {\isacharequal}\ {\isacharparenleft}{\isasymnot}\ qf{\isacharunderscore}form{\isadigit{2}}{\isacharunderscore}interp\ qf{\isacharunderscore}form\ vs{\isacharparenright}{\isachardoublequoteclose}\ {\isacharbar}\isanewline
        \ \ {\isachardoublequoteopen}qf{\isacharunderscore}form{\isadigit{2}}{\isacharunderscore}interp\ {\isacharparenleft}Conj\ qf{\isacharunderscore}form{\isadigit{1}}\ norm{\isacharunderscore}form{\isadigit{2}}{\isacharparenright}\ vs\ \isanewline
        \ \ \ \ {\isacharequal}\ {\isacharparenleft}qf{\isacharunderscore}form{\isadigit{2}}{\isacharunderscore}interp\ qf{\isacharunderscore}form{\isadigit{1}}\ vs\ {\isasymand}\ qf{\isacharunderscore}form{\isadigit{2}}{\isacharunderscore}interp\ norm{\isacharunderscore}form{\isadigit{2}}\ vs{\isacharparenright}{\isachardoublequoteclose}{\isacharbar}\isanewline
        \ \ {\isachardoublequoteopen}qf{\isacharunderscore}form{\isadigit{2}}{\isacharunderscore}interp\ {\isacharparenleft}Disj\ qf{\isacharunderscore}form{\isadigit{1}}\ qf{\isacharunderscore}form{\isadigit{2}}{\isacharparenright}\ vs\ \isanewline
        \ \ \ \ {\isacharequal}\ {\isacharparenleft}qf{\isacharunderscore}form{\isadigit{2}}{\isacharunderscore}interp\ qf{\isacharunderscore}form{\isadigit{1}}\ vs\ {\isasymor}\ qf{\isacharunderscore}form{\isadigit{2}}{\isacharunderscore}interp\ qf{\isacharunderscore}form{\isadigit{2}}\ vs{\isacharparenright}{\isachardoublequoteclose}{\isacharbar}\isanewline
        \ \ {\isachardoublequoteopen}qf{\isacharunderscore}form{\isadigit{2}}{\isacharunderscore}interp\ T\ vs\ {\isacharequal}\ True{\isachardoublequoteclose}{\isacharbar}\isanewline
        \ \ {\isachardoublequoteopen}qf{\isacharunderscore}form{\isadigit{2}}{\isacharunderscore}interp\ F\ vs\ {\isacharequal}\ False{\isachardoublequoteclose}
\end{isabelle}

\begin{isabelle}
        \isacommand{fun}\isamarkupfalse%
        \ norm{\isacharunderscore}form{\isadigit{2}}{\isacharunderscore}interp{\isacharcolon}{\isacharcolon}\ {\isachardoublequoteopen}norm{\isacharunderscore}form{\isadigit{2}}\ {\isasymRightarrow}real\ list\ {\isasymRightarrow}\ bool{\isachardoublequoteclose}\ \isakeyword{where}\isanewline
        \ \ {\isachardoublequoteopen}norm{\isacharunderscore}form{\isadigit{2}}{\isacharunderscore}interp\ {\isacharparenleft}QF\ qf{\isacharparenright}\ vs\ {\isacharequal}\ qf{\isacharunderscore}form{\isadigit{2}}{\isacharunderscore}interp\ qf\ vs{\isachardoublequoteclose}{\isacharbar}\isanewline
        \ \ {\isachardoublequoteopen}norm{\isacharunderscore}form{\isadigit{2}}{\isacharunderscore}interp\ {\isacharparenleft}ExQ\ norm{\isacharunderscore}form{\isacharparenright}\ vs\ \isanewline
        \ \ \ \ \ \ {\isacharequal}\ {\isacharparenleft}{\isasymexists}x{\isachardot}\ norm{\isacharunderscore}form{\isadigit{2}}{\isacharunderscore}interp\ norm{\isacharunderscore}form\ {\isacharparenleft}x{\isacharhash}vs{\isacharparenright}{\isacharparenright}{\isachardoublequoteclose}{\isacharbar}\isanewline
        \ \ {\isachardoublequoteopen}norm{\isacharunderscore}form{\isadigit{2}}{\isacharunderscore}interp\ {\isacharparenleft}AllQ\ norm{\isacharunderscore}form{\isacharparenright}\ vs\ \isanewline
        \ \ \ \ \ \ {\isacharequal}\ {\isacharparenleft}{\isasymforall}x{\isachardot}\ norm{\isacharunderscore}form{\isadigit{2}}{\isacharunderscore}interp\ norm{\isacharunderscore}form\ {\isacharparenleft}x{\isacharhash}vs{\isacharparenright}{\isacharparenright}{\isachardoublequoteclose}
\end{isabelle}

Given the definition of a (structured) datatype \isa{norm\_form2} and the corresponding interpretation function \isa{norm\_form2\_interp}, target formulas can now be parsed. For example, we can convert a univariate formula
\begin{isabelle}
 {\isachardoublequoteopen}{\isasymforall}x{\isacharcolon}{\isacharcolon}real{\isachardot}\ x\ {\isachargreater}\ {\isadigit{1}}{\isacharslash}{\isadigit{2}}\ {\isasymor}\ x\ {\isacharless}\ {\isadigit{1}}{\isachardoublequoteclose}
\end{isabelle}
into an equivalent form
\begin{isabelle}
        {\isachardoublequoteopen}norm{\isacharunderscore}form{\isadigit{2}}{\isacharunderscore}interp\isanewline
        \ \ \ {\isacharparenleft}AllQ\ {\isacharparenleft}QF\ {\isacharparenleft}Disj\ {\isacharparenleft}Pos\ {\isacharparenleft}Pol\ {\isacharbrackleft}{\isacharcolon}{\isacharminus}\ {\isadigit{1}}{\isacharcomma}\ {\isadigit{2}}{\isacharcolon}{\isacharbrackright}\ {\isadigit{0}}{\isacharparenright}{\isacharparenright}\ \isanewline
        \ \ \ \ \ \ \ \ \ \ \ \ \ \ \ \ \ \ \ {\isacharparenleft}Pos\ {\isacharparenleft}Pol\ {\isacharbrackleft}{\isacharcolon}{\isadigit{1}}{\isacharcomma}\ {\isacharminus}\ {\isadigit{1}}{\isacharcolon}{\isacharbrackright}\ {\isadigit{0}}{\isacharparenright}{\isacharparenright}\isanewline
        \ \ \ \ \ \ \ \ \ \ \ \ \ {\isacharparenright}{\isacharparenright}{\isacharparenright}\isanewline
        \ \ \ {\isacharbrackleft}{\isacharbrackright}{\isachardoublequoteclose}
\end{isabelle}
In particular, note
\begin{isabelle}
 qf{\isacharunderscore}form{\isadigit{2}}{\isacharunderscore}interp\ {\isacharparenleft}Pos\ {\isacharparenleft}Pol\ {\isacharbrackleft}{\isacharcolon}{\isacharminus}\ {\isadigit{1}}{\isacharcomma}\ {\isadigit{2}}{\isacharcolon}{\isacharbrackright}\ {\isadigit{0}}{\isacharparenright}{\isacharparenright}\ {\isacharbrackleft}x{\isacharbrackright}\
 \newline
  {\isacharequal}\ {\isacharparenleft}poly\ {\isacharbrackleft}{\isacharcolon}{\isacharminus}\ {\isadigit{1}}{\isacharcomma}\ {\isadigit{2}}{\isacharcolon}{\isacharbrackright}\ x\ {\isachargreater}\ {\isadigit{0}}{\isacharparenright}\
  \newline
   {\isacharequal}\ {\isacharparenleft}x\ {\isachargreater}\ {\isadigit{1}}{\isacharslash}{\isadigit{2}}{\isacharparenright}
\end{isabelle}
in which inequalities have been parsed into a polynomial sign determination problem.

On the contrary, a bivariate non-closed formula such as
\begin{isabelle}
        {\isachardoublequoteopen}{\isasymexists}x{\isacharcolon}{\isacharcolon}real{\isachardot}\ x\ {\isacharplus}\ y\ {\isachargreater}{\isadigit{0}}{\isachardoublequoteclose}
\end{isabelle}
will be converted into
\begin{isabelle}
        {\isachardoublequoteopen}norm{\isacharunderscore}form{\isadigit{2}}{\isacharunderscore}interp\isanewline
        \ \ {\isacharparenleft}ExQ\ {\isacharparenleft}QF\ {\isacharparenleft}Pos\isanewline
        \ \ \ \ \ \ \ \ {\isacharparenleft}Abnorm\isanewline
        \ \ \ \ \ \ \ \ \ \ {\isacharparenleft}Add\ {\isacharparenleft}Add\ {\isacharparenleft}Add\ {\isacharparenleft}C\ {\isadigit{0}}{\isacharparenright}\ {\isacharparenleft}Mul\ {\isacharparenleft}Var\ {\isadigit{0}}{\isacharparenright}\ {\isacharparenleft}Add\ {\isacharparenleft}C\ {\isadigit{1}}{\isacharparenright}\ {\isacharparenleft}Mul\ {\isacharparenleft}Var\ {\isadigit{0}}{\isacharparenright}\ {\isacharparenleft}C\ {\isadigit{0}}{\isacharparenright}{\isacharparenright}{\isacharparenright}{\isacharparenright}{\isacharparenright}\isanewline
        \ \ \ \ \ \ \ \ \ \ \ \ \ \ \ \ \ \ \ \ {\isacharparenleft}Add\ {\isacharparenleft}C\ {\isadigit{0}}{\isacharparenright}\ {\isacharparenleft}Mul\ {\isacharparenleft}Var\ {\isadigit{1}}{\isacharparenright}\ {\isacharparenleft}Add\ {\isacharparenleft}C\ {\isadigit{1}}{\isacharparenright}\ {\isacharparenleft}Mul\ {\isacharparenleft}Var\ {\isadigit{1}}{\isacharparenright}\ {\isacharparenleft}C\ {\isadigit{0}}{\isacharparenright}{\isacharparenright}{\isacharparenright}{\isacharparenright}{\isacharparenright}{\isacharparenright}\isanewline
        \ \ \ \ \ \ \ \ \ \ \ \ \ \ \ {\isacharparenleft}C\ {\isadigit{0}}{\isacharparenright}{\isacharparenright}{\isacharparenright}{\isacharparenright}{\isacharparenright}{\isacharparenright}\isanewline
        \ \ \ \ \ {\isacharbrackleft}y{\isacharbrackright}{\isachardoublequoteclose}
\end{isabelle}
where the \isa{Abnorm} constructor indicates that such formula is not supported by our current tactic.

\subsection{Existential Case}
To discharge a univariate existential formula is easy: we can computationally check if a certificate (i.e., a real algebraic number) returned by an external solver satisfies the quantifier-free part of the formula:

\begin{isabelle}
\isacommand{lemma}\isamarkupfalse%
\ ExQ{\isacharunderscore}intro{\isacharcolon}\isanewline
\ \ \isakeyword{fixes}\ x{\isacharcolon}{\isacharcolon}{\isachardoublequoteopen}alg{\isacharunderscore}float{\isachardoublequoteclose}\ \isakeyword{and}\ qf{\isacharunderscore}form{\isacharcolon}{\isacharcolon}qf{\isacharunderscore}form{\isadigit{2}}\isanewline
\ \ \isakeyword{assumes}\ {\isachardoublequoteopen}qf{\isacharunderscore}form{\isadigit{2}}{\isacharunderscore}interp\ qf{\isacharunderscore}form\ {\isacharbrackleft}of{\isacharunderscore}alg{\isacharunderscore}float\ x{\isacharbrackright}{\isachardoublequoteclose}\isanewline
\ \ \isakeyword{shows}\ {\isachardoublequoteopen}norm{\isacharunderscore}form{\isadigit{2}}{\isacharunderscore}interp\ {\isacharparenleft}ExQ\ {\isacharparenleft}QF\ qf{\isacharunderscore}form{\isacharparenright}{\isacharparenright}\ {\isacharbrackleft}{\isacharbrackright}{\isachardoublequoteclose}
\end{isabelle}
where \isa{x} of type \isa{alg\_float}
\begin{isabelle}
        \isacommand{datatype}\isamarkupfalse%
        \ alg{\isacharunderscore}float\ {\isacharequal}\ \isanewline
        \ \ \ \ \ Arep\ {\isachardoublequoteopen}int\ poly{\isachardoublequoteclose}\ float\ float\ %
        \isamarkupcmt{representation of a real algebraic number%
        }
        \isanewline
        \ \ \ {\isacharbar}\ Flt\ float\ \isamarkupcmt{a small optimization in case the number is dyadic rational %
        }
\end{isabelle}
 is a certificate that  is supposed to be instantiated by an external solver. The function \isa{of\_alg\_float} converts \isa{x} from \isa{alg\_float} to \isa{real}. In other words, to prove an existential formula:
\begin{isabelle}
        {\isachardoublequoteopen}norm{\isacharunderscore}form{\isadigit{2}}{\isacharunderscore}interp\ {\isacharparenleft}ExQ\ {\isacharparenleft}QF\ qf{\isacharunderscore}form{\isacharparenright}{\isacharparenright}\ {\isacharbrackleft}{\isacharbrackright}{\isachardoublequoteclose}
\end{isabelle}
we can computationally check the truth value of the quantifier-free part of the formula at \isa{x}:
\begin{isabelle}
        {\isachardoublequoteopen}qf{\isacharunderscore}form{\isadigit{2}}{\isacharunderscore}interp\ qf{\isacharunderscore}form\ {\isacharbrackleft}of{\isacharunderscore}alg{\isacharunderscore}float\ x{\isacharbrackright}{\isachardoublequoteclose}
\end{isabelle}
which is possible due to the sign determination procedure described in Sec. \ref{sec:deciding_sign}.

\subsection{Universal Case}

For the universal case, the core lemma is as follows:
\begin{isabelle}
        \isacommand{lemma}\isamarkupfalse%
        \ utilize{\isacharunderscore}samples{\isacharcolon}\isanewline
        \ \ \isakeyword{fixes}\ P{\isacharcolon}{\isacharcolon}{\isachardoublequoteopen}real\ {\isasymRightarrow}\ bool{\isachardoublequoteclose}\ \isakeyword{and}\ decomps{\isacharcolon}{\isacharcolon}{\isachardoublequoteopen}real\ set\ set{\isachardoublequoteclose}\ \isanewline
        \ \ \ \ \isakeyword{and}\ samples{\isacharcolon}{\isacharcolon}{\isachardoublequoteopen}real\ set{\isachardoublequoteclose}\ \isakeyword{and}\ f{\isacharcolon}{\isacharcolon}{\isachardoublequoteopen}real\ set\ {\isasymRightarrow}\ real{\isachardoublequoteclose}\isanewline
        \ \ \isakeyword{assumes}\ {\isachardoublequoteopen}{\isasymUnion}decomps\ {\isacharequal}\ {\isasymreal}{\isachardoublequoteclose}\isanewline
        \ \ \ \ \ \ \isakeyword{and}\ {\isachardoublequoteopen}{\isasymforall}d{\isasymin}decomps{\isachardot}\ {\isasymforall}x{\isadigit{1}}{\isasymin}d{\isachardot}{\isasymforall}x{\isadigit{2}}{\isasymin}d{\isachardot}\ P\ x{\isadigit{1}}\ {\isacharequal}\ P\ x{\isadigit{2}}{\isachardoublequoteclose}\isanewline
        \ \ \ \ \ \ \isakeyword{and}\ {\isachardoublequoteopen}{\isasymforall}d{\isasymin}decomps{\isachardot}\ f\ d{\isasymin}d{\isachardoublequoteclose}\ \isakeyword{and}\ {\isachardoublequoteopen}bij{\isacharunderscore}betw\ f\ decomps\ samples{\isachardoublequoteclose}\ \isanewline
        \ \ \isakeyword{shows}\ {\isachardoublequoteopen}{\isacharparenleft}{\isasymforall}x{\isachardot}\ P\ x{\isacharparenright}\ {\isacharequal}\ {\isacharparenleft}{\isasymforall}pt{\isasymin}samples{\isachardot}\ P\ pt{\isacharparenright}{\isachardoublequoteclose}
\end{isabelle}
where \isa{{bij{\isacharunderscore}betw\ f\ decomps\ samples}} states that
\isa{f::real set \isasymRightarrow\ real} is a bijective function between the
decomposition \isa{decomps::real set set} and the sample points
\isa{samples::real set}. Essentially, what the lemma \isa{utilize\_samples}
shows is that given a predicate \isa{P::real \isasymRightarrow\ bool}, an
unbounded universal formula \isa{{\isasymforall}x{\isachardot}\ P\ x} is
equivalent to a bounded one
\isa{{\isasymforall}pt{\isasymin}samples{\isachardot}\ P\ pt}, if the truth
value of \isa{P} is constant over each component of the decomposition:
\isa{{\isasymforall}d{\isasymin}decomps{\isachardot}\ {\isasymforall}x{\isadigit{1}}{\isasymin}d{\isachardot}{\isasymforall}x{\isadigit{2}}{\isasymin}d{\isachardot}\ P\ x{\isadigit{1}}\ {\isacharequal}\ P\ x{\isadigit{2}}}.

On top of the lemma \isa{utilize\_samples}, we similarly convert an unbounded
univariate real formula into a bounded one:
\begin{isabelle}
        \isacommand{lemma}\isamarkupfalse%
        \ allQ{\isacharunderscore}subst{\isacharcolon}\isanewline
        \ \ \isakeyword{fixes}\ root{\isacharunderscore}reps{\isacharcolon}{\isacharcolon}{\isachardoublequoteopen}alg{\isacharunderscore}float\ list{\isachardoublequoteclose}\ \isakeyword{and}\ pols{\isacharcolon}{\isacharcolon}{\isachardoublequoteopen}float\ poly\ set{\isachardoublequoteclose}\isanewline
        \ \ \ \ \isakeyword{and}\ qf{\isacharunderscore}form{\isacharcolon}{\isacharcolon}qf{\isacharunderscore}form{\isadigit{2}}\isanewline
        \ \ \isakeyword{defines}\ {\isachardoublequoteopen}samples{\isasymequiv}map\ of{\isacharunderscore}alg{\isacharunderscore}float\ {\isacharparenleft}mk{\isacharunderscore}samples\ root{\isacharunderscore}reps{\isacharparenright}{\isachardoublequoteclose}\isanewline
        \ \ \isakeyword{assumes}\ {\isachardoublequoteopen}Some\ pols\ {\isacharequal}\ extractPols\ qf{\isacharunderscore}form{\isachardoublequoteclose}\ \isanewline
        \ \ \ \ \ \ \isakeyword{and}\ {\isachardoublequoteopen}ordered{\isacharunderscore}reps\ root{\isacharunderscore}reps{\isachardoublequoteclose}\ \isanewline
        \ \ \ \ \ \ \isakeyword{and}\ {\isachardoublequoteopen}contain{\isacharunderscore}all{\isacharunderscore}roots\ root{\isacharunderscore}reps\ pols{\isachardoublequoteclose}\ \isanewline
        \ \ \ \ \ \ \isakeyword{and}\ {\isachardoublequoteopen}valid{\isacharunderscore}list\ root{\isacharunderscore}reps{\isachardoublequoteclose}\isanewline
        \ \ \isakeyword{shows}\ {\isachardoublequoteopen}norm{\isacharunderscore}form{\isadigit{2}}{\isacharunderscore}interp\ {\isacharparenleft}AllQ\ {\isacharparenleft}QF\ qf{\isacharunderscore}form{\isacharparenright}{\isacharparenright}\ vs\ \isanewline
        \ \ \ \ {\isacharequal}\ {\isacharparenleft}{\isasymforall}x\ {\isasymin}\ {\isacharparenleft}set\ samples{\isacharparenright}{\isachardot}\ norm{\isacharunderscore}form{\isadigit{2}}{\isacharunderscore}interp\ {\isacharparenleft}QF\ qf{\isacharunderscore}form{\isacharparenright}\ {\isacharparenleft}x{\isacharhash}vs{\isacharparenright}{\isacharparenright}{\isachardoublequoteclose}
\end{isabelle}
where
\begin{itemize}
        \item \isa{root\_reps::alg\_float list} is a certificate that
          should be instantiated by an external solver. More
          specifically, \isa{root\_reps} should be the representation
          of a list of real roots (in ascending order) of polynomials
          from the quantifier-free part of the target formula,
        \item
          \isa{map\ of{\isacharunderscore}alg{\isacharunderscore}float\ {\isacharparenleft}mk{\isacharunderscore}samples\ root{\isacharunderscore}reps{\isacharparenright}}
          constructs sample points from the representation of a list
          of roots,
        \item \isa{extractPols qf\_form} extracts polynomials from the
          quantifier-free part \isa{qf\_form},
        \item \isa{ordered\_reps root\_reps} and \isa{valid\_list
          root\_reps} together ensure that the representation of roots
          are valid and those roots are in ascending order,
        \item \isa{contain\_all\_roots roots\_reps pols} checks if
          \isa{root\_reps} is a representation of all real roots of
          the polynomials \isa{pols}. Specifically, by Sturm's
          theorem, the number of total distinct real roots of each
          \isa{p \isasymin\ pols} can be computed, which can be then
          compared with the number of \isa{r \isasymin\ root\_reps}
          that \isa{p(r)=0}.
\end{itemize}
Most importantly, all assumptions of the lemma \isa{allQ\_subst} and its right-hand side
\begin{isabelle}
        {\isacharparenleft}{\isasymforall}x\ {\isasymin}\ {\isacharparenleft}set\ samples{\isacharparenright}{\isachardot}\ norm{\isacharunderscore}form{\isadigit{2}}{\isacharunderscore}interp\ {\isacharparenleft}QF\ qf{\isacharunderscore}form{\isacharparenright}\ {\isacharparenleft}x{\isacharhash}vs{\isacharparenright}{\isacharparenright}
\end{isabelle}
can be computationally checked, through which we can prove an unbounded univariate universal formula:
\isa{norm{\isacharunderscore}form{\isadigit{2}}{\isacharunderscore}interp\ {\isacharparenleft}AllQ\ {\isacharparenleft}QF\ qf{\isacharunderscore}form{\isacharparenright}{\isacharparenright}\ vs}.

\section{Linking to an External Solver} \label{sec:linking_to_external}

Certificates for both existential and universal cases can be produced by any
program performing univariate CAD. For now, we implement the program on top of
Mathematica. More specifically, the universal certificates are constructed by
the Mathematica command \emph{SemialgebraicComponentInstances}, which gives
sample points in each connected component of a semialgebraic set. The
existential certificates are constructed by the command \emph{FindInstance},
which incorporates powerful numerical methods to accelerate the search
for real algebraic sample points.

Also, it may be worth mentioning that after a certificate has been found, our
tactic will record it (as a string) so that repeating the proof no longer
requires the external solver. This is much like the sums-of-squares tactic
\cite{daumas2009verified}.

In general, the certificate-based design grants us much flexibility: We can
easily switch to a more efficient external solver without modifying existing
formal proofs. In fact, we were first using an implementation of univariate CAD
built within MetiTarski, which turned out to be not very efficient, and we
simply switched to the current one based on Mathematica. In the future, we plan
to experiment with other open-source CAD implementations such as Z3 and QEPCAD
to provide more options with external solvers.

\section{Experiments and Related Work} \label{sec:related_work}
\begin{figure}

        \begin{align*}
        \mathrm{ex1:}\quad&\forall x.\, \neg (x \geq -9 \land x<10 \land x^4 >0 ) \lor x^{12} > 0 \\
        \\%
        \mathrm{ex2:}\quad &\forall x.\, \neg ((x-2)^2 (-x+4)>0 \land x^2(x-3)^2 \geq 0 \\
        &\land x-1 \geq 0 \land -(x-2)^2 + 1 >0) \lor (-(x-\frac{11}{12}))^3 (x-\frac{41}{10})^3 \geq 0 \\
        %
        %\mathrm{ex3:}\quad&\exists x.\, (x-2)^2(-x+4)>0 \wedge x^2(x-3)^2 \geq 0 \wedge x-1 \geq 0 \\&\wedge - (x-3)^2 + 1 >0 \wedge -(x-\frac{11}{12})^3(x-\frac{41}{10})^3 < \frac{1}{10} \\
        \\%
        \mathrm{ex3:}\quad&\exists x.\, x^5 - x -1 =0 \land x^{12} + \frac{425}{23} x^{11} -\frac{228}{23} x^{10} - 2 x^8 - \frac{896}{23} x^7 - \frac{394}{23} x^6 + \\
        & \frac{456}{23} x^5 + x^4 + \frac{471}{23} x^3 + \frac{645}{23} x^2 - \frac{31}{23} x - \frac{228}{23} =0 \land x^3 + 22 x^2 - 31 \geq 0 \\
        & \land x^{22} - \frac{234}{567} x^{20} - 419 x^{10} + 1948 >0 \\
        \\ %
        \mathrm{ex4:}\quad& \forall x.\, x>0\lor \frac{20}{9} x^3+\frac{5}{9} x^2-\frac{61}{9} x>-4\lor 1\leq x\lor x\leq 0\lor \frac{10}{9}x^2-\frac{19}{9} x\leq -1 \\
        & \lor \frac{1}{18}x^3+\frac{31}{45} x^2-\frac{13}{9}x\leq -\frac{7}{10}\lor \frac{20}{9} x^3+\frac{5 }{9}x^2-\frac{61}{9} x\leq -4\\
        \\%
        \mathrm{ex5:}\quad& \forall x.\,  -\frac{x^3}{3}-\frac{10}{3} x^2-\frac{5}{6} x>0\lor \frac{1}{3}x^3+\frac{10}{3} x^2+\frac{5}{6} x>0\lor 1\leq x\lor x\leq 0 \\ & \lor \frac{10}{9} x^2-\frac{19}{9} x\leq -1\lor \frac{1}{18}x^3+\frac{31}{45} x^2-\frac{13}{9} x\leq -\frac{7}{10} \\ & \lor \frac{14}{15} x^3-\frac{64}{15} x^2-\frac{101}{30} x\leq -\frac{11}{5}\lor \frac{20}{9} x^3+\frac{5}{9} x^2-\frac{61}{9} x\leq -4 \\
        \\%
        \mathrm{ex6:}\quad& \exists x.\, -70 x^6-\frac{2052}{5} x^5-\frac{4329}{5}  x^4-\frac{5409}{10} x^3-\frac{267}{2}  x^2-\frac{51}{10} x>-\frac{7}{10}\land \frac{49}{162} x^9+\frac{49}{3} x^8 \\ & +\frac{175}{18} x^7+\frac{115774}{405} x^6+\frac{77743}{135} x^5-\frac{57328}{135} x^4-\frac{135853}{810} x^3-\frac{71681}{270} x^2-\frac{10327}{270} x>-\frac{721}{90} \\ & \land \frac{7}{27} x^8+\frac{280}{27} x^7-\frac{595}{54} x^6+\frac{18964}{135} x^5+\frac{2698}{135} x^4-\frac{24217}{270} x^3-\frac{251}{6} x^2-\frac{2981}{90} x>-\frac{206}{45} \\ & \land \frac{7}{54} x^7+\frac{112}{27} x^6+\frac{329}{90} x^5+\frac{2672}{135} x^4-\frac{7933}{270} x^3+\frac{169}{18} x^2-\frac{799}{90} x>-\frac{103}{90}\land \frac{7}{27} x^8+\frac{280}{27} x^7 \\ & +\frac{935}{54} x^6+\frac{7264}{135} x^5+\frac{11323}{135} x^4-\frac{12217}{270} x^3-\frac{701}{6} x^2-\frac{781}{90} x>-\frac{77}{15}\land \frac{2}{9} x^7+\frac{52}{9} x^6-\frac{17}{6} x^5 \\ & +\frac{2353}{90} x^4+\frac{307}{45} x^3-\frac{811}{30} x^2-\frac{361}{30} x>-\frac{44}{15}\land \frac{1}{9}x^6+2 x^5+\frac{2}{15} x^4+\frac{41}{90} x^3-\frac{2}{15} x^2 \\ & -\frac{33}{10} x>-\frac{11}{15}\land \frac{49}{162} x^8+\frac{1540}{81} x^7+\frac{1109}{27} x^6+\frac{23483}{810} x^5+\frac{65378}{405} x^4-\frac{11549}{270} x^3-\frac{70225}{324} x^2 \\ & -\frac{1339}{405} x>-\frac{721}{60}\land \frac{7}{27} x^7+\frac{203}{18} x^6-\frac{52}{9} x^5+\frac{7753}{270} x^4+\frac{5191}{180} x^3-\frac{2263}{45} x^2-\frac{10741}{540} x>-\frac{103}{15} \\ & \land \frac{2}{9} x^6+\frac{59}{9} x^5-\frac{493}{36} x^4+\frac{2113}{90} x^3-\frac{811}{180} x^2-\frac{1481}{90} x>-\frac{22}{5}\land \frac{1}{9}x^5+\frac{17}{9} x^4-\frac{257}{60} x^3+\frac{563}{90} x^2 \\ &-\frac{913}{180} x>-\frac{11}{10}\land \frac{20}{9} x^4-\frac{5}{2} x^3+\frac{10}{3} x^2-\frac{91}{18} x>-2\land \frac{10}{9} x^3-\frac{25}{18} x^2-\frac{2}{9} x>-\frac{1}{2}\land \frac{20}{9} x^3 \\ & +\frac{5}{9} x^2-\frac{61}{9} x>-4\land 1>x\land x>0\land \frac{10}{9} x^2-\frac{19}{9} x>-1\land \frac{1}{18}x^3+\frac{31}{45} x^2-\frac{13}{9} x>-\frac{7}{10}\land \frac{1}{9}x^4 \\ & +\frac{34}{15} x^3-\frac{53}{30} x^2-\frac{253}{90} x>-\frac{11}{5}\land \frac{2}{9} x^5+\frac{82}{9} x^4+\frac{86}{15} x^3-\frac{2051}{90} x^2-\frac{97}{90} x>-\frac{44}{5}\land \frac{8}{81} x^8 \\ & +\frac{931}{81} x^7+\frac{3113}{27} x^6-\frac{289811}{1620} x^5+\frac{264373}{810} x^4+\frac{30583}{270} x^3-\frac{298609}{810} x^2-\frac{93307}{1620} x>-\frac{193}{5}\\ & \land \frac{7}{27} x^7+\frac{38}{3} x^6+\frac{28}{9} x^5-\frac{2686}{135} x^4+\frac{6397}{60} x^3-\frac{9151}{90} x^2-\frac{4741}{540} x>-\frac{77}{10}\\
        \end{align*}
        \phantomcaption
\end{figure}
\begin{figure}
        \ContinuedFloat
        \begin{align*}
        \mathrm{ex7:}\quad& \forall x.\, x<-1\lor 0>x\lor \frac{1}{8} x^7+\frac{1207}{35} x^6+\frac{7083}{10} x^5+4983 x^4+\frac{64405}{4} x^3+26169 x^2 \\ & +\frac{41613 }{2}x>-6435 \lor 35 x^{12}+22461058620 x^2+11821609800 x\leq 46204 x^{11} \\ & +5263834 x^{10}+144537452 x^9+1758662439 x^8+10317027768 x^7+31842714428 x^6 \\ & +54212099480 x^5+45938678170 x^4+4171407240 x^3\lor  x\leq 0\lor 753 x^{10}+58568 x^9 \\ & +938908 x^8+6857016 x^7+27930066 x^6+68338600 x^5+102560612 x^4+92372280 x^3 \\ & +45805760 x^2+9609600 x\leq 0\lor 10 x^{11}+1101329460 x^2+788107320 x\leq 9179 x^{10} \\ & +1061504 x^9+24397102 x^8+240283734 x^7+1063536663 x^6+2362290448 x^5 \\ & +2625491260 x^4+782617220 x^3\lor 5 x^{10}+81290790 x^2+90935460 x\leq 2828 x^9 \\ &+356071 x^8+6846880 x^7+51834563 x^6+161529144 x^5+237512625 x^4 \\ & +125595120 x^3\lor 207 x^9+11237 x^8+138652 x^7+794964 x^6+2505504 x^5+4581220 x^4 \\ & +4837448 x^3+2735040 x^2+640640 x\leq 0\lor 5 x^8\leq 608 x^7+10261 x^6+63520 x^5 \\ & +192458 x^4+303324 x^3+238560 x^2+73920 x\lor 98 x^8+3514 x^7+32711 x^6+142928 x^5 \\ & +332962 x^4+424284 x^3+278880 x^2+73920 x\leq 0\lor x\leq -1 \\
        \end{align*}

        \center
        \begin{tabular}{@{}cccc@{}} \toprule
                & \multicolumn{3}{c}{Time (s)} \\ \cmidrule(r){2-4}
                Formula & univ\_rcf (Isabelle)  & univ\_rcf\_cert (Isabelle) & tarski (PVS)\\ \midrule
                ex1 & 0.9  & 0.3  & 2.0  \\
                ex2 & 1.4  & 0.6  & 6.8  \\
                ex3 & 1.6  & 0.7  & 13.0  \\
                ex4 & 1.3 & 0.5  & 20.1  \\
                ex5 & 1.6 & 0.6  & 315.7  \\
                ex6 & 5.6 & 3.9  &   timeout \\
                ex7 & 38.4 & 34.9  & timeout  \\
                \bottomrule
                \multicolumn{4}{l}{\footnotesize{Note: timeout indicates failure to terminate within 24 hours}}
        \end{tabular}
        \caption{Comparison between our tactic in Isabelle and the
          tarski strategy in PVS: univ\_rcf includes
          certificate searching and checking, while univ\_rcf\_cert
          includes only checking} \label{fig:experiments}
\end{figure}

The most relevant work is the recent {\tt tarski} strategy by Narkawicz et al.\
\cite{narkawicz2015formally} in PVS. Both their work and ours rely on a
formal proof of the Sturm-Tarski theorem (which they call Tarski's theorem) and
handle roughly the same class of problems\footnote{In fact, their tactic does not handle arbitrary boolean expressions like ours, but we believe this should not be too hard to overcome.} (i.e., first-order  univariate formulas over
reals). There are two main differences between their work and ours:
\begin{itemize}
\item Their procedure resembles Tarski's original quantifier
  elimination \cite[Chapter~2]{real_alg_geo2006} and Cyril Cohen's quantifier
  elimination procedure in Coq \cite[Chapter~12]{cohen_phd} by making use of
  both the Sturm-Tarski theorem and matrices. In contrast, our tactic is based
  on CAD and real algebraic numbers (instead of matrices).
\item Their procedure is entirely built within PVS, while ours sceptically makes
  use of efficient external programs to generate certificates.
\end{itemize}

To compare both tactics empirically, we have conducted experiments on several
typical examples from their paper\footnote{\url{http://shemesh.larc.nasa.gov/people/cam/Tarski/}} and the MetiTarski project\footnote{\url{http://www.cl.cam.ac.uk/~gp351/cicm2012/}} \cite{passmore2012real}.
The experiments
are run on a desktop with an Intel Core 2 Quad Q9400 (quad core, 2.66 GHz) CPU
and 8 gigabytes RAM. Results of the experiments are illustrated in
Fig.~\ref{fig:experiments}, where our \isa{univ\_rcf} tactic includes both
certificate searching and checking process, while the \isa{univ\_rcf\_cert} does
the checking part only (when repeating a proof with certificates already
recorded as a string).

In general, the experiments indicate that our tactic outperforms the
{\tt tarski} strategy in PVS. Particularly, the advantage of our
tactic becomes greater as the problems become more complex, which can
be attributed to the fact that our tactic has much better worst-case
computational complexity (polynomial vs. exponential in the number of
polynomials).

In the case of general multivariate problems, the CAD procedure is
doubly exponential while Tarski's quantifier elimination procedure is
non-elementary in the number of variables
\cite[Chapter~11]{real_alg_geo2006}). When limited to univariate
problems, the CAD procedure degenerates to root isolation and sign
determination on a set of univariate polynomials, which is of
polynomial complexity in the number of polynomials and their degree
bound \cite[Chapter~10]{real_alg_geo2006}). In comparison, Tarski's
quantifier elimination procedure, even when limited to univariate
problems, is still exponential in the number of polynomials
\cite{cohen2012formal}.

In addition, it is worth noting that as the problems become more
complex (e.g., ex6 and ex7 in Fig.~\ref{fig:experiments}), certificate
checking becomes the bottleneck factor of our tactic (especially for
universal problems). This indicates that, despite the fact that
certificate searching is much harder than certificate checking, the
Mathematica implementation is still much more efficient than our
verified certificate-checking procedure. This leaves much room for
future optimisations.

Our work has also been greatly inspired by Cyril Cohen's PhD thesis
\cite{cohen_phd}, within which a quantifier elimination procedure has been built
upon the Sturm-Tarski theorem and real algebraic numbers formalised within the
Coq theorem prover. However, our goals and approaches are very different.

Cohen's work is part of a large project that has formalised the Feit-Thompson
theorem (odd order theorem) in Coq \cite{gonthier2013machine}, and focuses more
on theoretical developments than we do. For example, they proved the
Sturm-Tarski theorem to construct an RCF quantifier elimination procedure in the
spirit of Tarski's original method, which has important theoretical properties
but is not practical as a proof procedure. Moreover, he has formalised
arithmetic on real algebraic numbers and shown that they form a real closed
field via resultants. We have not formalised resultants at all. Our sign
determination algorithm uses the Sturm-Tarski theorem, which is significantly
more efficient in practice than using resultants. On the other hand, as it was
unnecessary for our proof procedure, we have not proved in Isabelle that the
real algebraic numbers form a real closed field. In general, compared to his
work, ours stresses the practical side over the theoretical. Fundamentally, we
want to build procedures to solve non-trivial problems in practice.

Decision procedures based on Sturm's theorem have been implemented in Isabelle
and PVS before \cite{Eberl,NM2014NASA}. Their core idea is to count the number
of real roots within a certain (bounded or unbounded) interval. Generally, they
can only handle formulas involving a single polynomial, so they are not complete
for first-order formulas (unlike our tactic and the {\tt tarski} strategy in
PVS).

Assia Mahboubi \cite{mahboubi2007implementing} has implemented the executable
part of a general CAD procedure in Coq, but as far as we
know, the correctness proof for her implementation is still ongoing. This is also
one of the reasons for us to choose the certificate-based approach rather than
directly verifying an implementation.

There are other methods to handle nonlinear polynomial problems in theorem
provers, such as sums of squares \cite{daumas2009verified}, which is good for
multivariate universal problems but is not applicable when the existential
quantifier arises, and interval arithmetic
\cite{holzl2009proving,solovyev2013formal}, which is very efficient for some
cases but is not complete. These methods and ours should be used in a
complementary way.

\section{Discussion and Applications}\label{sec:discussion_metitarski}

One of our driving motivations is the integration of MetiTarski with Isabelle.
%In this section, we present some applications of our new \isa{univ\_rcf} tactic in verifying real algebraic reasoning used during MetiTarski proofs.
%
MetiTarski \cite{metitarski-jar} is a first-order theorem prover for real number
inequalities involving transcendental functions such as $\sin$, $\tan$ and
$\exp$. It can automatically prove formulas like
\[
\forall x \in (0,1.25).\,  \tan(x)^2 \leq 1.75 \times 10^{-7} + \tan(1)\tan(x^2)
\]
\[
\forall x > 0.\, \frac{1-e^{-2 x}}{2x(1-e^{-x})^2} -\frac{1}{x^2} \leq \frac{1}{12}
\]
\[
\forall x \in (0,1).\, 1.914 \frac{\sqrt{1+x}-\sqrt{1-x}}{4+\sqrt{1+x}+\sqrt{1-x}} \leq 0.01+\frac{x}{2+\sqrt{1-x^2}}.
\]

The main idea behind MetiTarski is to approximate transcendental functions by
polynomial or rational function bounds, and then solve the formula by a
combination of a resolution theorem proving and an external Real Closed Field
(RCF) decision procedure (QEPCAD, Mathematica or Z3). MetiTarski is a version of
Joe Hurd's Metis prover \cite{hurd2007metis}, modified to include arithmetic
simplification and integration with RCF decision procedures, along with many
other refinements.

Applications of MetiTarski include verification problems arising in air traffic
control \cite{denman2011towards} and analogue circuit designs
\cite{denman2009formal}. As some of the applications are safety critical, it is
natural to consider to integrate MetiTarski with an existing interactive theorem
prover, whose internal logic can be used to ensure the correctness of
MetiTarski's proofs. Besides, the automation provided by MetiTarski is generally
useful to interactive theorem provers.

MetiTarski has been integrated with the PVS theorem prover \cite{owre1992pvs} as
a trusted oracle \cite{denman2014automated}. The authors state that the
automation introduced by MetiTarski for closing sequents containing real-valued
functions considerably outperforms existing tactics in PVS\@. However, this
tactic should not be used in a certification environment, where external oracles
are not allowed.

Our eventual goal is to integrate MetiTarski into the Isabelle/HOL theorem
prover. Isabelle can verify purely logical inferences (in fact, it contains an
internal copy of the Metis theorem prover), and the third author has just
formalised most of the bounds of transcendental functions used by MetiTarski
\cite{Special_Function_Bounds}. The primary remaining hurdle is the RCF decision
procedure, and the work presented here is the first step towards it.

Finally, let us say a bit about how our work might be generalised to
multivariate problems. In doing so, we plan to continue our certificate-based
approach, as we are unlikely to implement a verified internal CAD procedure
comparable in efficiency to a state-of-the-art implementation. It is still not
obvious to us where the clear separation between \emph{search} and
\emph{verification} should be in the multivariate case, but we have already made
some progress:
\begin{itemize}
\item The bivariate sign determination procedure based on recursive application
  of the Sturm-Tarski theorem described in our previous work \cite{li2016modular}
  can be easily generalised to a multivariate one (i.e., a procedure to decide the
  sign of a multivariate polynomial at real algebraic points), which can be then
  used to efficiently certify purely existential multivariate formulas over
  reals.
\item Our recent formalisation of Cauchy's residue theorem
  \cite{residue_itp_2016} can be used to certify a key theorem used in general
  CAD: that the complex roots of a polynomial continuously depend on its coefficients.
\end{itemize}

\section{Conclusion} \label{sec:conclusion}

We have described our work of building a procedure for
first-order univariate polynomial problems in Isabelle/HOL. Compared to existing
tactics among proof assistants, noticeable features of our tactic are
\begin{itemize}
\item It is based on univariate cylindrical algebraic decomposition (CAD).
\item It sceptically integrates efficient external solvers in a
  certificate-based way, so that its soundness solely depends on Isabelle's
  logic (and code generation machinery) rather than the external solvers.
\end{itemize}
This is made possible by certificate-based approaches to real root isolation and
sign-determination for evaluating polynomials at real algebraic points.
As much of the novelty in our work is motivated by practical efficiency
considerations, we have performed experiments comparing our procedure with another
real algebraic proof procedure, the {\tt tarski} method in PVS.
By making use of efficient external solvers, our procedure is shown to
empirically outperform this other method by substantial margins.
We believe this adds further impetus to the certificate-based
methods for a wide variety of formal proof procedures.

Certificate-based methods can be compared on the basis of how much
mathematics and computation are required both to find and check their
certificates.
For example, to convert a Positivstellensatz certificate into a
HOL-Light proof of a universal theorem, Harrison's sums-of-squares
tactic only requires simple sign-based reasoning and rational
arithmetic, while in our case, we need more mathematics (e.g., real
algebraic numbers and the Sturm-Tarski theorem) and more computation
(especially for the universal case).
A good certificate design needs to balance the difficulty of the
formalisation effort and verified computation required to check the
certificates with the efficiency improvements offered by offloading
the construction of the certificates to high-performance external
tools.

\paragraph*{Acknowledgements.}

We thank Florian Haftmann for helping with code generation for our procedure. We are also grateful to the anonymous referees for their constructive suggestions.

\bibliographystyle{spmpsci}

\bibliography{real_closed_field}

\begin{thebibliography}{10}
\providecommand{\url}[1]{{#1}}
\providecommand{\urlprefix}{URL }
\expandafter\ifx\csname urlstyle\endcsname\relax
  \providecommand{\doi}[1]{DOI~\discretionary{}{}{}#1}\else
  \providecommand{\doi}{DOI~\discretionary{}{}{}\begingroup
  \urlstyle{rm}\Url}\fi

\bibitem{metitarski-jar}
Akbarpour, B., Paulson, L.: {MetiTarski}: An automatic theorem prover for
  real-valued special functions.
\newblock Journal of Automated Reasoning \textbf{44}(3), 175--205 (2010)

\bibitem{real_alg_geo2006}
Basu, S., Pollack, R., Roy, M.F.: Algorithms in Real Algebraic Geometry
  (Algorithms and Computation in Mathematics).
\newblock Springer-Verlag New York, Inc., Secaucus, NJ, USA (2006)

\bibitem{brown2003qepcad}
Brown, C.W.: {QEPCAD B}: a program for computing with semi-algebraic sets using
  {CAD}s.
\newblock ACM SIGSAM Bulletin \textbf{37}(4), 97--108 (2003)

\bibitem{chaieb2008automated}
Chaieb, A., et~al.: Automated methods for formal proofs in simple arithmetics
  and algebra.
\newblock Diss., Technische Universit{\"a}t, M{\"u}nchen  (2008)

\bibitem{cheng2007complete}
Cheng, J.S., Gao, X.S., Yap, C.K.: Complete numerical isolation of real zeros
  in zero-dimensional triangular systems.
\newblock In: Proceedings of the 2007 international symposium on Symbolic and
  algebraic computation, pp. 92--99. ACM (2007)

\bibitem{cohen_phd}
Cohen, C.: {Formalized algebraic numbers: construction and first-order theory}.
\newblock Ph.D. thesis, {\'E}cole polytechnique (2012)

\bibitem{cohen2012formal}
Cohen, C., Mahboubi, A., et~al.: Formal proofs in real algebraic geometry: from
  ordered fields to quantifier elimination.
\newblock Logical Methods in Computer Science \textbf{8}(1: 02), 1--40 (2012)

\bibitem{collins1976quantifier}
Collins, G.E.: Quantifier elimination for real closed fields by cylindrical
  algebraic decomposition: a synopsis.
\newblock ACM SIGSAM Bulletin \textbf{10}(1), 10--12 (1976)

\bibitem{de2008z3}
De~Moura, L., Bj{\o}rner, N.: Z3: An efficient smt solver.
\newblock In: Tools and Algorithms for the Construction and Analysis of
  Systems, pp. 337--340. Springer (2008)

\bibitem{de2013computation}
De~Moura, L., Passmore, G.O.: Computation in real closed infinitesimal and
  transcendental extensions of the rationals.
\newblock In: International Conference on Automated Deduction, pp. 178--192.
  Springer Berlin Heidelberg (2013)

\bibitem{denman2009formal}
Denman, W., Akbarpour, B., Tahar, S., Zaki, M.H., Paulson, L.C.: Formal
  verification of analog designs using {MetiTarski}.
\newblock In: Formal Methods in Computer-Aided Design, 2009. FMCAD 2009, pp.
  93--100. IEEE (2009)

\bibitem{denman2014automated}
Denman, W., Mu{\~n}oz, C.: Automated real proving in {PVS} via {MetiTarski}.
\newblock In: FM 2014: Formal Methods, pp. 194--199. Springer (2014)

\bibitem{denman2011towards}
Denman, W., Zaki, M.H., Tahar, S., Rodrigues, L.: Towards flight control
  verification using automated theorem proving.
\newblock In: NASA Formal Methods, pp. 89--100. Springer (2011)

\bibitem{Eberl}
Eberl, M.: A decision procedure for univariate real polynomials in
  {I}sabelle/{HOL}.
\newblock In: Proceedings of the 2015 Conference on Certified Programs and
  Proofs, CPP '15, pp. 75--83. ACM, New York, NY, USA (2015).
\newblock \doi{10.1145/2676724.2693166}.
\newblock \urlprefix\url{http://doi.acm.org/10.1145/2676724.2693166}

\bibitem{gonthier2013machine}
Gonthier, G., Asperti, A., Avigad, J., Bertot, Y., Cohen, C., Garillot, F.,
  Le~Roux, S., Mahboubi, A., O¡¯Connor, R., Biha, S.O., et~al.: A
  machine-checked proof of the odd order theorem.
\newblock In: Interactive Theorem Proving, pp. 163--179. Springer (2013)

\bibitem{haftmann2010code}
Haftmann, F., Nipkow, T.: Code generation via higher-order rewrite systems.
\newblock In: International Symposium on Functional and Logic Programming, pp.
  103--117. Springer (2010)

\bibitem{daumas2009verified}
Harrison, J.: Verifying nonlinear real formulas via sums of squares.
\newblock In: K.~Schneider, J.~Brandt (eds.) Proceedings of the 20th
  International Conference on Theorem Proving in Higher Order Logics, TPHOLs
  2007, \emph{Lecture Notes in Computer Science}, vol. 4732, pp. 102--118.
  Springer-Verlag, Kaiserslautern, Germany (2007)

\bibitem{holzl2009proving}
H{\"o}lzl, J.: Proving inequalities over reals with computation in
  {Isabelle/HOL}.
\newblock In: International Workshop on Programming Languages for Mechanized
  Mathematics Systems, pp. 38--45 (2009)

\bibitem{hurd2007metis}
Hurd, J.: Metis first order prover.
\newblock Website at http://gilith. com/software/metis  (2007)

\bibitem{residue_itp_2016}
Li, W., Paulson, L.C.: A formal proof of {C}auchy's residue theorem.
\newblock In: ITP 2016: Seventh International Conference on Interactive Theorem
  Proving, p. to appear (2016)

\bibitem{li2016modular}
Li, W., Paulson, L.C.: A modular, efficient formalisation of real algebraic
  numbers.
\newblock In: Proceedings of the 5th ACM SIGPLAN Conference on Certified
  Programs and Proofs, pp. 66--75. ACM (2016)

\bibitem{mahboubi2007implementing}
Mahboubi, A.: Implementing the cylindrical algebraic decomposition within the
  {Coq} system.
\newblock Mathematical Structures in Computer Science \textbf{17}(1), 99--127
  (2007)

\bibitem{Mishra}
Mishra, B.: Algorithmic Algebra.
\newblock Springer-Verlag New York, Inc., New York, NY, USA (1993)

\bibitem{pvs_bernstein}
Mu{\~{n}}oz, C., Narkawicz, A.: Formalization of {B}ernstein polynomials and
  applications to global optimization.
\newblock Journal of Automated Reasoning \textbf{51}(2), 151--196 (2013).
\newblock \doi{10.1007/s10817-012-9256-3}.
\newblock \urlprefix\url{http://dx.doi.org/10.1007/s10817-012-9256-3}

\bibitem{narkawicz2015formally}
Narkawicz, A., Munoz, C., Dutle, A.: Formally-verified decision procedures for
  univariate polynomial computation based on {S}turm's and {T}arski's theorems.
\newblock Journal of Automated Reasoning \textbf{54}(4), 285--326 (2015)

\bibitem{NM2014NASA}
Narkawicz, A.J., Mu{\~{n}}oz, C.A.: A formally-verified decision procedure for
  univariate polynomial computation based on {S}turm's theorem.
\newblock Technical Memorandum NASA/TM-2014-218548, NASA, Langley Research
  Center, Hampton VA 23681-2199, USA (2014)

\bibitem{isa-tutorial}
Nipkow, T., Paulson, L.C., Wenzel, M.: Isabelle/HOL: A Proof Assistant for
  Higher-Order Logic.
\newblock Springer (2002)

\bibitem{owre1992pvs}
Owre, S., Rushby, J.M., Shankar, N.: {PVS}: A prototype verification system.
\newblock In: International Conference on Automated Deduction, pp. 748--752.
  Springer (1992)

\bibitem{passmore2012real}
Passmore, G.O., Paulson, L.C., De~Moura, L.: Real algebraic strategies for
  metitarski proofs.
\newblock In: International Conference on Intelligent Computer Mathematics, pp.
  358--370. Springer (2012)

\bibitem{Special_Function_Bounds}
Paulson, L.C.: Real-valued special functions: Upper and lower bounds.
\newblock Archive of Formal Proofs  (2014)

\bibitem{paulson2010three}
Paulson, L.C., Blanchette, J.C.: Three years of experience with {S}ledgehammer,
  a practical link between automatic and interactive theorem provers.
\newblock IWIL-2010 \textbf{1} (2010)

\bibitem{rahman2002analytic}
Rahman, Q., Schmeisser, G.: Analytic Theory of Polynomials.
\newblock London Mathematical Society monographs. Clarendon Press (2002).
\newblock \urlprefix\url{https://books.google.co.uk/books?id=FzFEEVO3PXYC}

\bibitem{sagraloff2010general}
Sagraloff, M.: A general approach to isolating roots of a bitstream polynomial.
\newblock Mathematics in Computer Science \textbf{4}(4), 481--506 (2010)

\bibitem{solovyev2013formal}
Solovyev, A., Hales, T.C.: Formal verification of nonlinear inequalities with
  taylor interval approximations.
\newblock In: NASA Formal Methods, pp. 383--397. Springer (2013)

\bibitem{strzebonski2006cylindrical}
Strzebo{\'n}ski, A.W.: Cylindrical algebraic decomposition using validated
  numerics.
\newblock Journal of Symbolic Computation \textbf{41}(9), 1021--1038 (2006)

\bibitem{Algebraic_Numbers_AFP}
Thiemann, R., Yamada, A.: Algebraic numbers in {Isabelle/HOL}.
\newblock Archive of Formal Proofs  (2015).
\newblock \url{http://isa-afp.org/entries/Algebraic_Numbers.shtml}, Formal
  proof development

\end{thebibliography}

\end{document}